\documentclass[usenatbib]{mnras}
\usepackage{ulem,amsmath,amssymb}
\usepackage{graphicx,subfigure}
\usepackage{enumitem,multirow}
\usepackage[dvipsnames]{xcolor}
\usepackage{color}
\begin{document}
\title[Optical Depths in Arp 220]{Breaking the Radio -- Gamma-Ray Connection in Arp 220}

\author[T. M. Yoast-Hull \& N. Murray]{Tova M. Yoast-Hull\thanks{E-mail: yoasthull@cita.utoronto.edu} and Norman Murray\thanks{Canadian Research Chair in Astrophysics} \\
Canadian Institute for Theoretical Astrophysics, University of Toronto, ON, M5S 3H8, Canada}

\maketitle

\begin{abstract}
Recent analyses of the $\gamma$-ray spectrum from the ultra-luminous infrared galaxy Arp 220 have revealed a discrepancy in the cosmic ray energy injection rates derived from the $\gamma$-rays versus the radio emission.  While the observed radio emission is consistent with the star formation rate inferred from infrared observations, a significantly higher cosmic ray population is necessary to accurately model the measured $\gamma$-ray flux.  To resolve this discrepancy between the radio and $\gamma$-ray observations, we find that we must increase the cosmic ray energy injection rate and account for an infrared optical depth greater than unity.  Raising the energy injection rate naturally raises the total $\gamma$-ray flux but also raises the radio flux unless there is also an increase in the energy loss rate for cosmic ray leptons.  A optically thick medium results in an increase in energy losses via inverse Compton for cosmic ray leptons and preserves agreement with submillimeter, millimeter, and infrared wavelength observations.
\end{abstract}


\section{Introduction} \label{sec:intro}

Over the past decade, eight star-forming galaxies (mostly giant spirals) have been detected in $\gamma$-rays by \textit{Fermi}, including M31 (NGC~0224), NGC~0253, NGC~1068, NGC~2146, M82 (NGC~3034), NGC~4945, Arp~220, and Circinus \citep{Abdo10a,Abdo10,Lenain10,Hayashida13,Tang14,Peng16,Griffin16}.  Observations of these galaxies show a tentative correlation between their $\gamma$-ray and far-infrared (FIR) luminosities \citep{Ackermann12,Rojas16}, indicating a physical relationship between these two quantities in star-forming galaxies as predicted by \citet{Thompson07}.  Recent simulations show further evidence of a link between the FIR emission and $\gamma$-rays from galaxies via their star formation rate \citep[SFR; e.g.,][]{Pohl14,Pfrommer17}.

The essential physical processes underlying this correlation are tied to massive stars, which dominate the diffuse emission in star-forming galaxies.  Massive stars radiate predominately in the ultraviolet; much of this emission is absorbed by the surrounding interstellar medium (ISM) and re-radiated in the infrared.  At the ends of their lives, massive stars explode as supernovae, and the resulting remnants accelerate electrons and protons to relativistic energies \citep[e.g.,][]{deJong85,Helou85}. Cosmic ray electrons interacting with galactic magnetic fields produce non-thermal synchrotron emission, while cosmic ray protons collide with the ISM to produce neutral pions that subsequently decay into $\gamma$-rays.  

In our current understanding, the correlation between radio and FIR emission seen in star-forming galaxies requires two additional conditions to hold.  First, either star-forming galaxies must be cosmic ray electron calorimeters, meaning that cosmic ray electrons are confined to the galaxy and radiate the majority of their energy within their injection volume \citep{Volk89}, or both cosmic ray electrons and ultraviolet radiation escape from star-forming galaxies in proportional amounts \citep{Lacki10}.  Secondly, in most star forming galaxies, energy losses via synchrotron must be larger than losses via inverse Compton for cosmic ray electrons, implying that the energy density in magnetic fields must be slightly larger than the energy density in radiation for star-forming galaxies \citep[$U_{B} \gtrsim U_{\rm rad}$;][]{Lisenfeld96a,Thompson06}.

While star-forming galaxies are generally believed to be electron calorimeters, most normal star-forming galaxies are not proton calorimeters \citep[e.g.,][]{Pfrommer17}.  In quiescent systems, the diffusion time-scale for cosmic ray protons is significantly shorter than their energy loss time-scale, and thus, cosmic ray protons preferentially escape into the halo.  Additionally, these are systems in which primary cosmic ray electrons are likely dominant over secondary cosmic ray electrons and positrons produced in interactions of cosmic ray protons with the ISM.

In contrast, intensely star-forming systems (such as Arp 220) are at least partial proton calorimeters \citep[e.g.,][]{Lacki11,YoastHull13,Eichmann16} and secondary cosmic rays dominate over primary cosmic ray electrons \citep[e.g.,][]{Lacki13b,Peretti18}.  As a result, in these systems, we would expect both the non-thermal radio and the $\gamma$-ray emission to be proportional to the population of cosmic ray protons.  Thus, for starburst systems in which secondary cosmic rays dominate, there may be a constant ratio between the $\gamma$-ray and radio fluxes, regardless of whether the galaxy is a complete proton calorimeter.

\begin{figure}
 \subfigure{
  \includegraphics[width=0.90\linewidth]{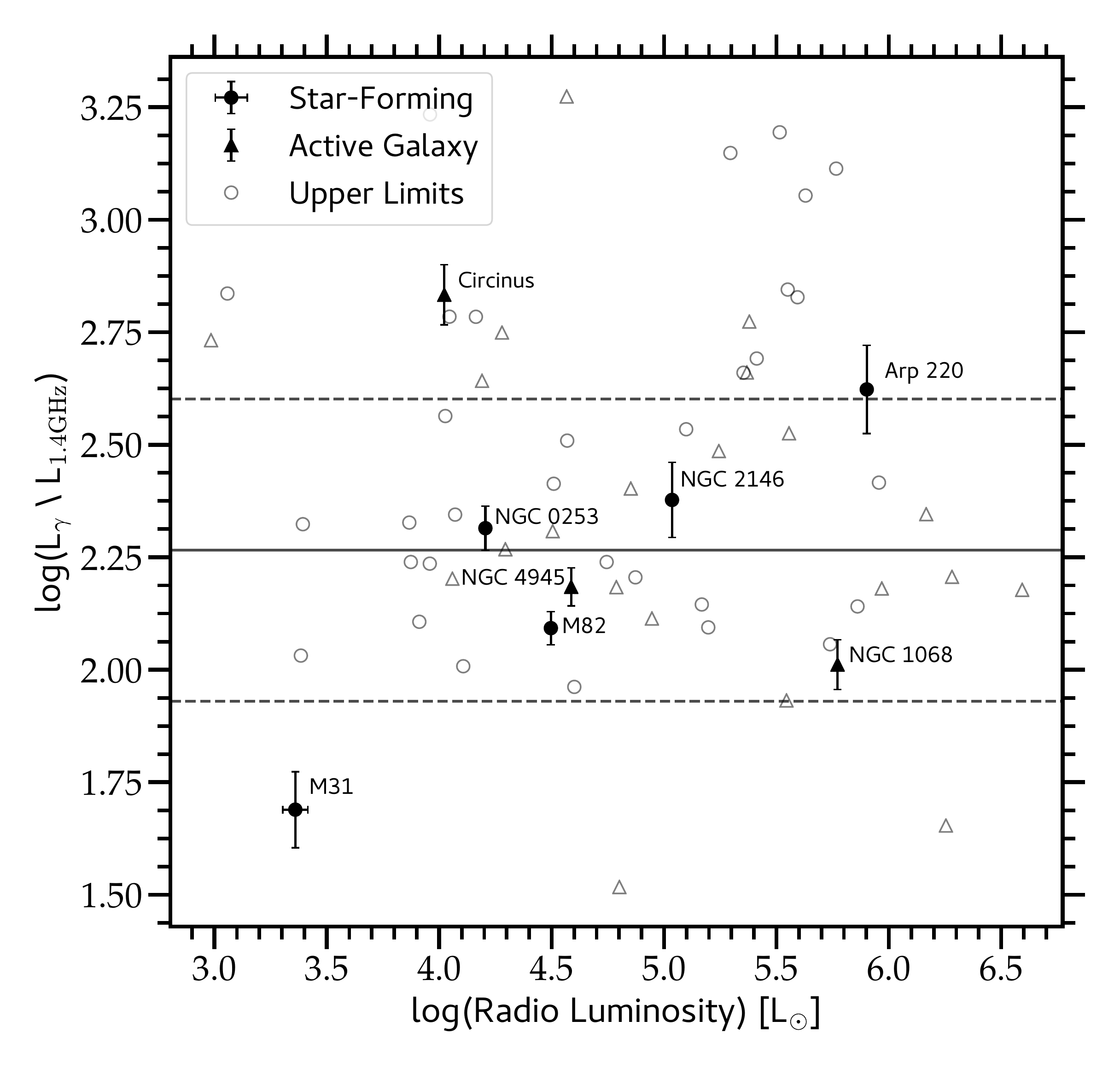}}
\caption{The logarithm of the ratio of $\gamma$-ray to radio luminosity, as a function of logarithmic radio luminosity, is shown for nearby star-forming galaxies.  The mean ratio is 2.266 with a standard deviation of 0.335, depicted here as solid and dashed lines respectively.  Local galaxy M31 and the Seyfert galaxy Circinus are clear outliers, while Arp 220 is a marginal outlier.  Galaxies with $\gamma$-ray detections by \textit{Fermi} are shown with filled black symbols and $\gamma$-ray flux upper limits are denoted with open gray symbols.  See Appendix~\ref{sec:appa} for details on the data used.}
\label{fig:lumratio}
\end{figure}

Observations clearly show that strong correlations exist between the FIR, radio, and $\gamma$-ray emission in star-forming galaxies \citep[e.g.,][]{Ackermann12,Schoneberg13,Rojas16}, see Fig.~\ref{fig:fluxflux} in Appendix~\ref{sec:appa}.  Examining the ratio of $\gamma$-ray flux to radio fluxes in star-forming galaxies detected with \textit{Fermi}, we find tentative evidence for a constant $\gamma$-ray to radio flux ratio in starbursts galaxies, see Fig.~\ref{fig:lumratio}.  The ratio for normal star-forming galaxies is lower than that for starbursts, as expected due to the dominance of primary cosmic ray electrons and the lack of proton calorimetry in quiescent galaxies.  In contrast, the ratio of $\gamma$-ray to radio flux for the Seyfert galaxy Circinus is significantly higher than the starburst ratio, likely due to the dominance of the central active galactic nucleus \citep[AGN; ][]{Wojaczynski15}.

\begin{figure}
 \subfigure{
  \includegraphics[width=0.90\linewidth]{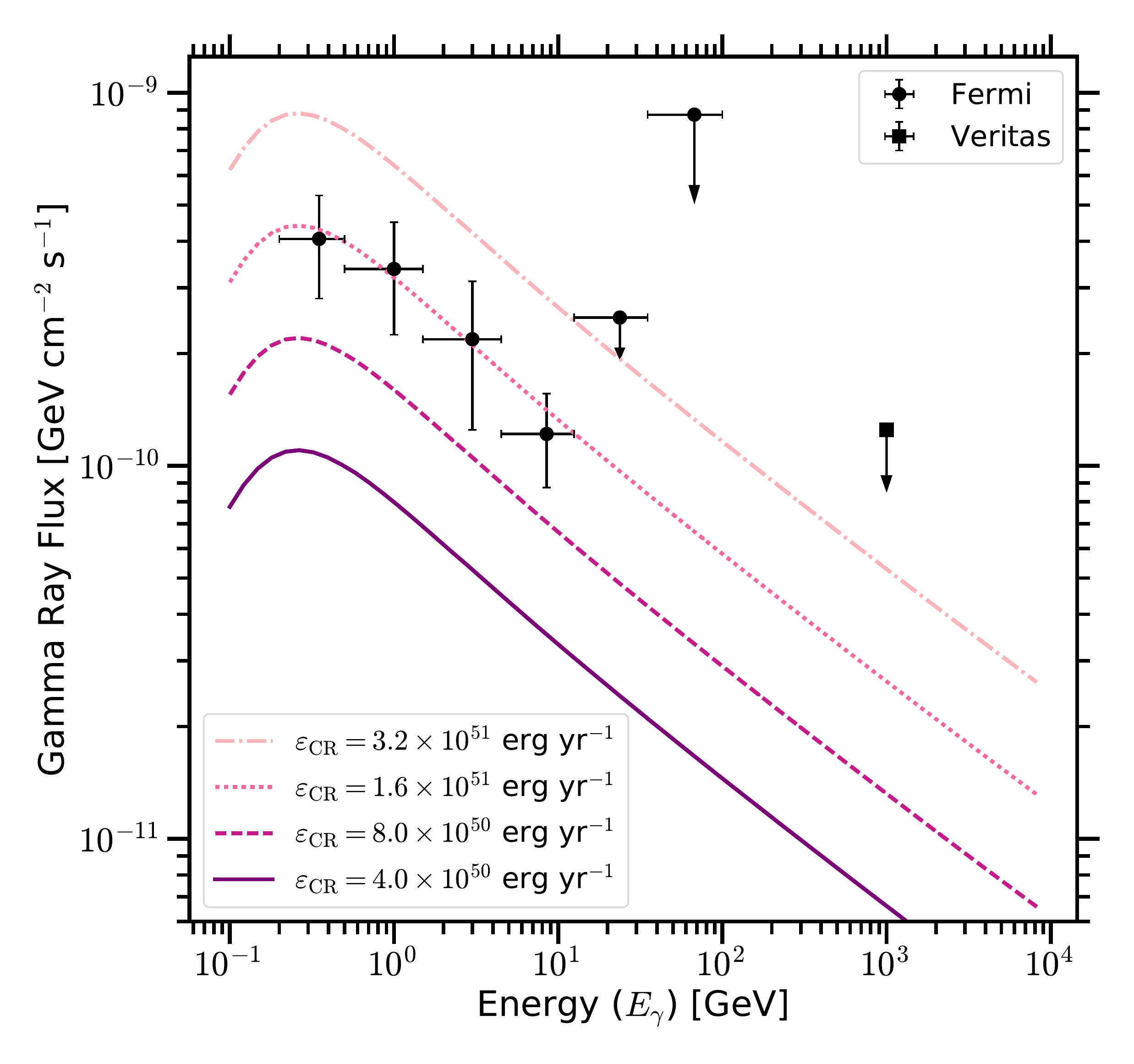}}
 \subfigure{
  \includegraphics[width=0.90\linewidth]{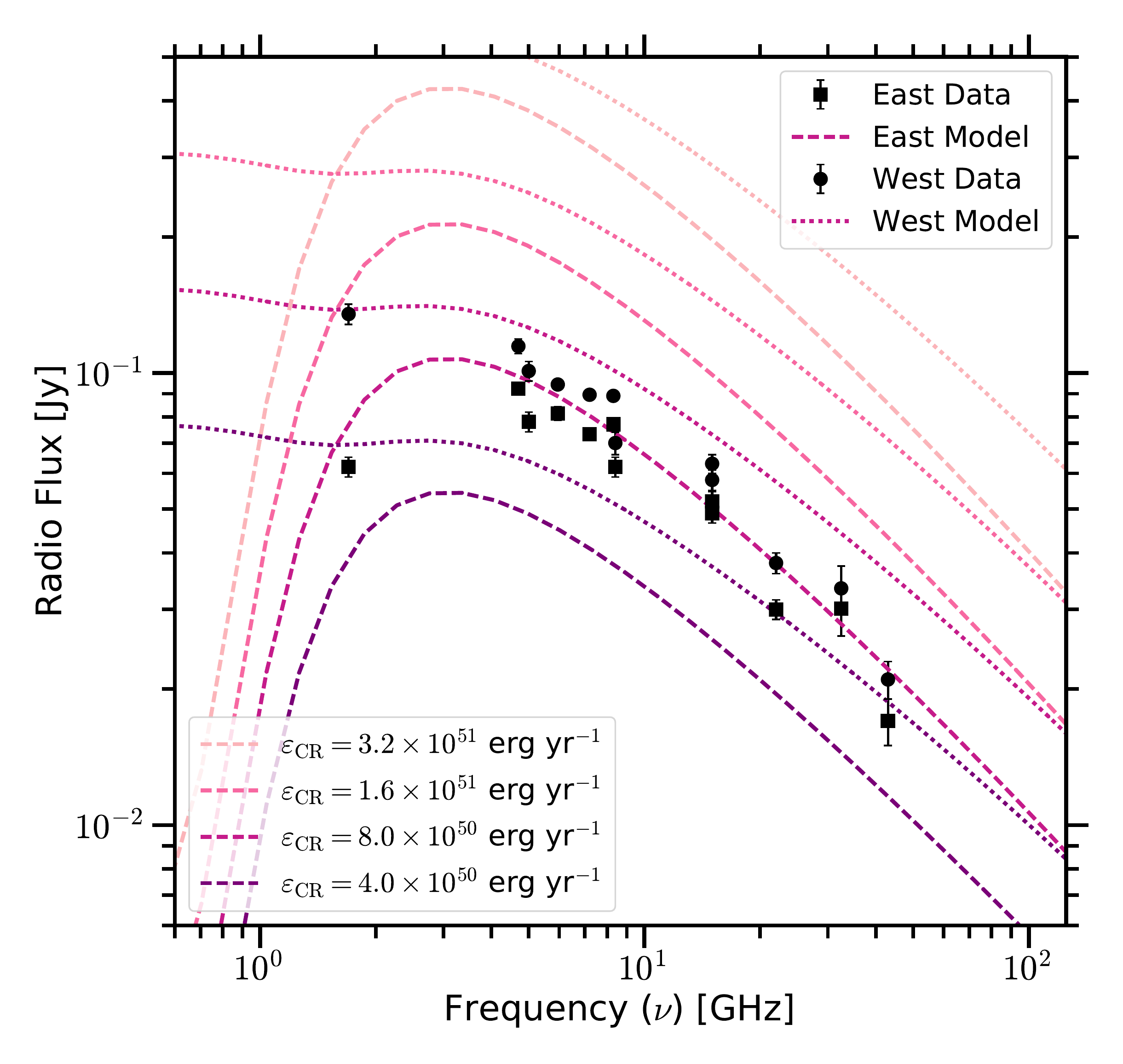}}
\caption{$\gamma$-ray and radio spectra are shown for varying cosmic ray energy injection rates. For the $\gamma$-rays, energy injection rates larger than $10^{51}$ erg~yr$^{-1}$ are necessary; however, energy injection rates lower than $10^{51}$ erg~yr$^{-1}$, are necessary to fit the radio data.  Data for the $\gamma$-rays are from \citet{Peng16,Fleischhack15}, and data for the radio are from \citet{Downes98,Rodriguez05,Williams10,Barcos15}.}
\label{fig:spectra}
\end{figure}

Fig.~\ref{fig:lumratio} also shows Arp 220, the closest ultra-luminous infrared galaxy (ULIRG), as slightly more than a standard deviation away from the mean $\gamma$-ray to radio luminosity ratio for starbursts.  In investigating Arp 220 in more detail, modeling of the $\gamma$-ray and radio spectra reveal a larger cosmic ray energy injection rate is necessary to reproduce the observed $\gamma$-ray flux than is necessary for the radio flux, see Fig.~\ref{fig:spectra}.  Of course, the magnitude of this discrepancy between the inferred energy injection rates depends on assumptions about the cosmic ray populations in the nuclei of Arp 220 \citep{YoastHull17a}.

\begin{figure*}
\includegraphics[width=0.9\linewidth]{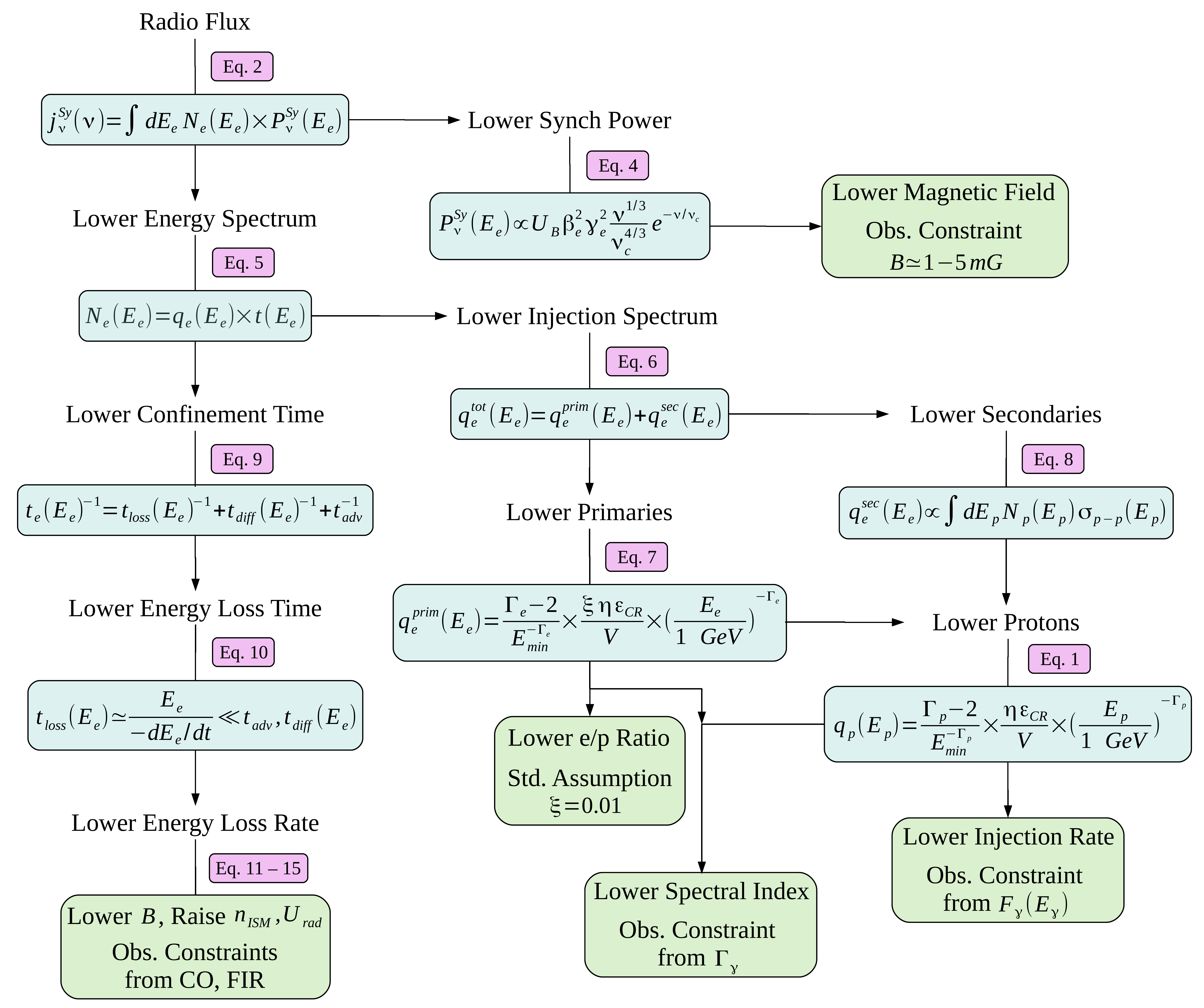}
\caption{Flow chart depicting the possible paths for lowering the radio flux.}
\label{fig:schematic}
\end{figure*}

To examine these assumptions in detail, we begin by defining the injection spectrum for cosmic ray protons such that
\begin{equation}
q_{p}(E_{p}) = \frac{\Gamma_{p} - 2}{E_{\rm min}^{-\Gamma_{p}}} \times \frac{\eta \varepsilon_{\rm CR}} {V} \times \left( \frac{E_{p}}{1 ~ \rm GeV} \right)^{-\Gamma_{p}}, \label{eqn:qp}
\end{equation}
where $\Gamma_{p}$ is the cosmic ray (injection) spectral index, $\varepsilon_{\rm CR}$ is the cosmic ray energy injection rate, and $\eta = 0.1$ is the acceleration efficiency of cosmic ray protons.  The critical factor here, besides the energy injection rate (see Fig.~\ref{fig:schematic}), is the cosmic ray proton spectral index: as the spectral index steepens, the energy injection rate required to supply a set $\gamma$-ray flux increases (because the factor $1/E_{\rm min}^{\Gamma_{p}}$ decreases).

Thus, for flatter spectral indices ($\Gamma_{p} \sim 2.1-2.2$), there exist models where the radio and $\gamma$-ray fluxes imply cosmic ray energy injection rates which are consistent with each other \cite[e.g.,][]{Torres04,Lacki11,Lacki13b}, and for steeper spectral indices ($\Gamma_{p} \sim 2.3-2.5$), the energy injection rates derived from the $\gamma$-ray and radio fluxes are inconsistent with each other \citep[e.g.,][]{YoastHull17a}.  As the observed $\gamma$-ray spectrum is best fit by spectral indices of $\Gamma_{\gamma} = 2.35 \pm 0.16$ \citep{Peng16}, both sets of models are consistent with the observations within the statistical uncertainty on the $\gamma$-ray spectral index\footnote{It should be noted here that $\Gamma_{p} \sim \Gamma_{\gamma}$, as the cross section for neutral pion production depends on energy such that $\sigma_{p-p}(E_{p}) \propto ln E_{p}$ \citep[e.g.,][]{Kelner06}.}.  Throughout the rest of this paper, we will be concentrating on solutions to the tension between the radio and $\gamma$-ray observations for steeper spectral indices.

In this paper, we examine possible mechanisms which can resolve the discrepancy between the observed radio and $\gamma$-ray fluxes from Arp 220.  Using our previously developed models for starburst galaxies \citep{YoastHull13,YoastHull14a,YoastHull15}, we explore how changes in energy injection rate, infrared optical depth, and magnetic field strength affect the cosmic ray electron populations and the resulting non-thermal emission, see Fig.~\ref{fig:schematic}.  We find that accounting for an infrared optical depth greater than unity in the models, as required by submillimeter and millimeter wavelength observations, easily resolves the tension between the energy injection rates derived from the radio and $\gamma$-ray spectra, respectively.

This paper is structured as follows: In Section~\ref{sec:theory}, we lay out the equations relating to the cosmic ray spectra and radio emission.  In Section~\ref{sec:constraints}, we discuss observational constraints on the physical quantities that are input into the models.  In Section~\ref{sec:results}, we describe how our models are constructed and tested, and we report our results.  In Section~\ref{sec:discussion}, we compare our results with observed and other quantities.  Our conclusions are summarized in Section~\ref{sec:summary}.

\section{Theoretical Framework} \label{sec:theory}

To resolve the difference between the cosmic ray energy injection rates inferred from $\gamma$-ray and radio spectra, we must lower the radio flux while maintaining or raising the $\gamma$-ray flux for a given cosmic ray energy injection rate.  Unfortunately, we cannot simply attribute the radio and $\gamma$-ray fluxes to different cosmic ray populations.

The primary hadronic $\gamma$-ray emission mechanism is neutral pion decay.  In the process of producing neutral pions, cosmic ray protons also create charged pions which subsequently decay into secondary electrons and positrons, along with various neutrinos.  Hence, the initial interactions leading to hadronic $\gamma$-rays also lead to radiation from secondary leptons and thus radio emission.  Therefore, whether the $\gamma$-rays are hadronic or leptonic in origin, there should also be a substantial amount of corresponding radio emission.

To explore all the variables affecting radio emission, we break down the various equations governing the radio synchrotron emissivity in Fig~\ref{fig:schematic}.  Inputs affecting the total synchrotron flux include: the magnetic field strength ($B$), the average ISM gas density ($n_{\rm ISM}$), the radiation field energy density ($U_{\rm rad}$), the electron-to-proton (number) ratio ($\xi$), the cosmic ray injection spectral index ($\Gamma_{e}$), and the cosmic ray energy injection rate ($\varepsilon_{\rm CR}$).

Both the radio and $\gamma$-ray spectra provide constraints on the cosmic ray spectral index, and the $\gamma$-ray flux, in particular, provides a lower limit on the the cosmic ray energy injection rate as discussed in Section~\ref{sec:intro}, limiting our avenues of altering the radio flux.  Additionally, lowering the electron-to-proton (number) ratio has limited effect due to the presence of secondary electrons and positrons.  Further observational constraints on the magnetic field strength, ISM density, and radiation field energy density are discussed in Section~\ref{sec:constraints}.

\subsection{Cosmic Ray Leptons} \label{sec:leptons}

Non-thermal emission dominates the radio spectrum at GHz frequencies, and so, we concentrate on the synchrotron emissivity,
\begin{equation}
j_{\nu}^{\rm Sy} = \int_{E_{\rm min}}^{\infty} dE_{e} N_{e}(E_{e}) P_{\nu}^{\rm Sy}(\gamma_{e}),\label{eqn:radio}
\end{equation}
where $N_{e}(E_{e})$ is the total leptonic energy spectrum. $P_{\nu}^{\rm Sy}$, the full synchrotron radiative power for a single electron, is given by \citep{Ginzburg69,Blumenthal70,RL79,Schlick02,Dermer09,Bottcher12,Ghisellini13}
\begin{equation}
P_{\nu}^{\rm Sy}(\gamma_{e},\psi) = \frac{\sqrt{3} q_{e}^{3} B}{m_{e} c^{2}} \sin(\psi) F(x).
\end{equation}
Here, $\psi$ is the pitch angle, and $F(x) = x \int_{x}^{\infty} d\xi K_{5/3}(\xi)$.  $K_{5/3}(\xi)$ is a modified Bessel function of the second kind, and $x = \nu / \nu_{\psi}$ where $\nu_{\psi} = 3 q_{e} B / 4 \pi m_{e} c \times \gamma_{e}^{2} \sin (\psi)$.  Assuming an isotropic pitch angle distribution and adopting the asymptotic approximation by \citet{Bottcher12}, the synchrotron power is given by
\begin{equation}
P_{\nu}^{\rm Sy} (\gamma_{e}) = \frac{32 \pi c}{9 \Gamma(4/3)} \left( \frac{q_{e}^{2}}{m_{e} c^{2}} \right)^{2} U_{B} \beta_{e}^{2} \gamma_{e}^{2} \frac{\nu^{1/3}}{\nu_{c}^{4/3}} e^{- \nu / \nu_{c}},\label{eqn:sypower}
\end{equation}
where $\Gamma(4/3)$ is the Gamma function and  the critical frequency $\nu_{c} = 3 q_{e} \gamma_{e}^{2} B / 4 \pi m_{e} c$.

\begin{figure}
 \subfigure{
  \includegraphics[width=0.90\linewidth]{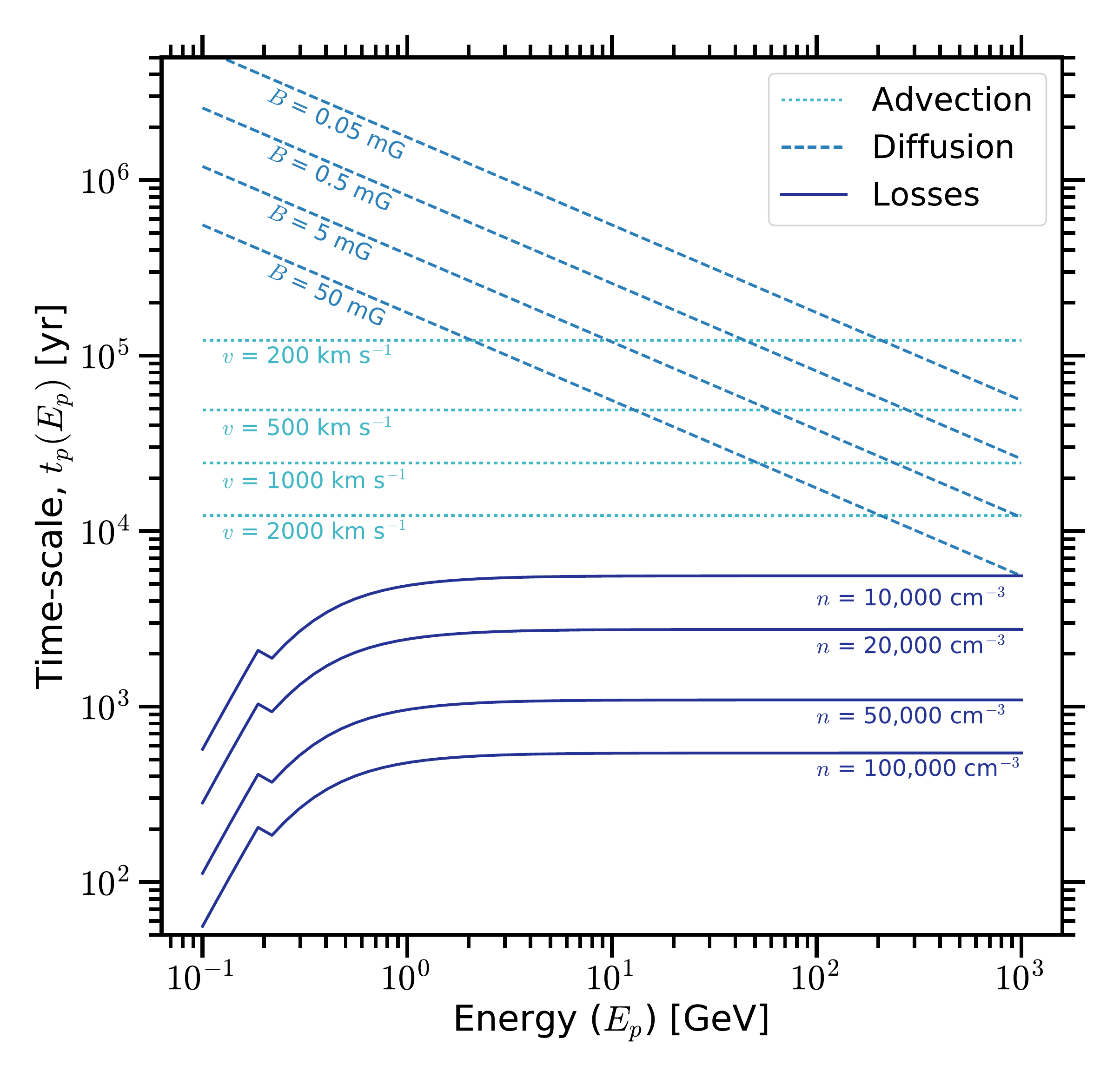}}
\caption{Confinement time-scales for cosmic ray protons are shown, including the diffusion time-scale (dashed lines), advection time-scale (dotted lines), and the energy-loss time-scale (solid lines).  The energy-loss time-scale for protons is dominated by the pion production time-scale, which is dependent on the average ISM gas density.  For the ISM densities found in Arp 220, the energy-loss time-scale is significantly lower than either the diffusion or advection time-scales.}
\label{fig:timescales}
\end{figure}

The energy spectrum for cosmic ray leptons is given by the product of the cosmic ray injection spectrum, $q(E)$, and the cosmic ray confinement time-scale, $t(E)$, see Fig.~\ref{fig:timescales}.
\begin{equation}
N(E) = q(E) \times t(E).\label{eqn:spectrum}
\end{equation}
The cosmic ray injection spectrum, $q(E)$, is assumed to be a power-law, proportional to the energy injection rate and inversely proportional to the volume of the injection region, as seen in the previous section.

The cosmic ray lepton spectrum includes both primary electrons, accelerated alongside the primary protons, and secondary electrons and positrons, the decay products of charged pions produced during collision of cosmic ray protons with the surrounding ISM.  Thus, the total cosmic ray lepton injection spectrum is given by
\begin{equation}
q_{e}^{\rm tot}(E_{e}) = q_{e}^{\rm prim}(E_{e}) + q_{e}^{\rm sec}(E_{e}),\label{eqn:qlepton}
\end{equation}
where the injection spectrum of the primary electrons is given by 
\begin{equation}
q_{e}^{\rm prim} = \frac{\Gamma_{e} - 2}{E_{\rm min}^{-\Gamma_{e}}} \times \frac{\xi \eta \varepsilon_{\rm CR}} {V} \times \left( \frac{E_{e}}{1 ~ \rm GeV} \right)^{-\Gamma_{e}}. \label{eqn:qprim}
\end{equation}
Here, $\xi$ is the electron-to-proton (number) ratio, which is usually taken to be $\xi = 0.01$ for $E_{e} = E_{p} = 1$~GeV.  For simplicity, we will assume that the spectral indices for the injection spectra for the cosmic ray protons and electrons are the same, $\Gamma_{p} = \Gamma_{e}$.  Then, the injection spectrum for secondary cosmic rays depends directly on the cosmic ray proton energy spectrum such that \citep[e.g.,][]{Schlick02,Kelner06}
\begin{equation}
q_{e}^{\rm sec}(E_{e}) \propto \int dE_{p} N_{p}(E_{p}) \sigma_{\rm p-p}(E_{p}).\label{eqn:qsec}
\end{equation}
\begin{figure}
 \subfigure{
  \includegraphics[width=0.90\linewidth]{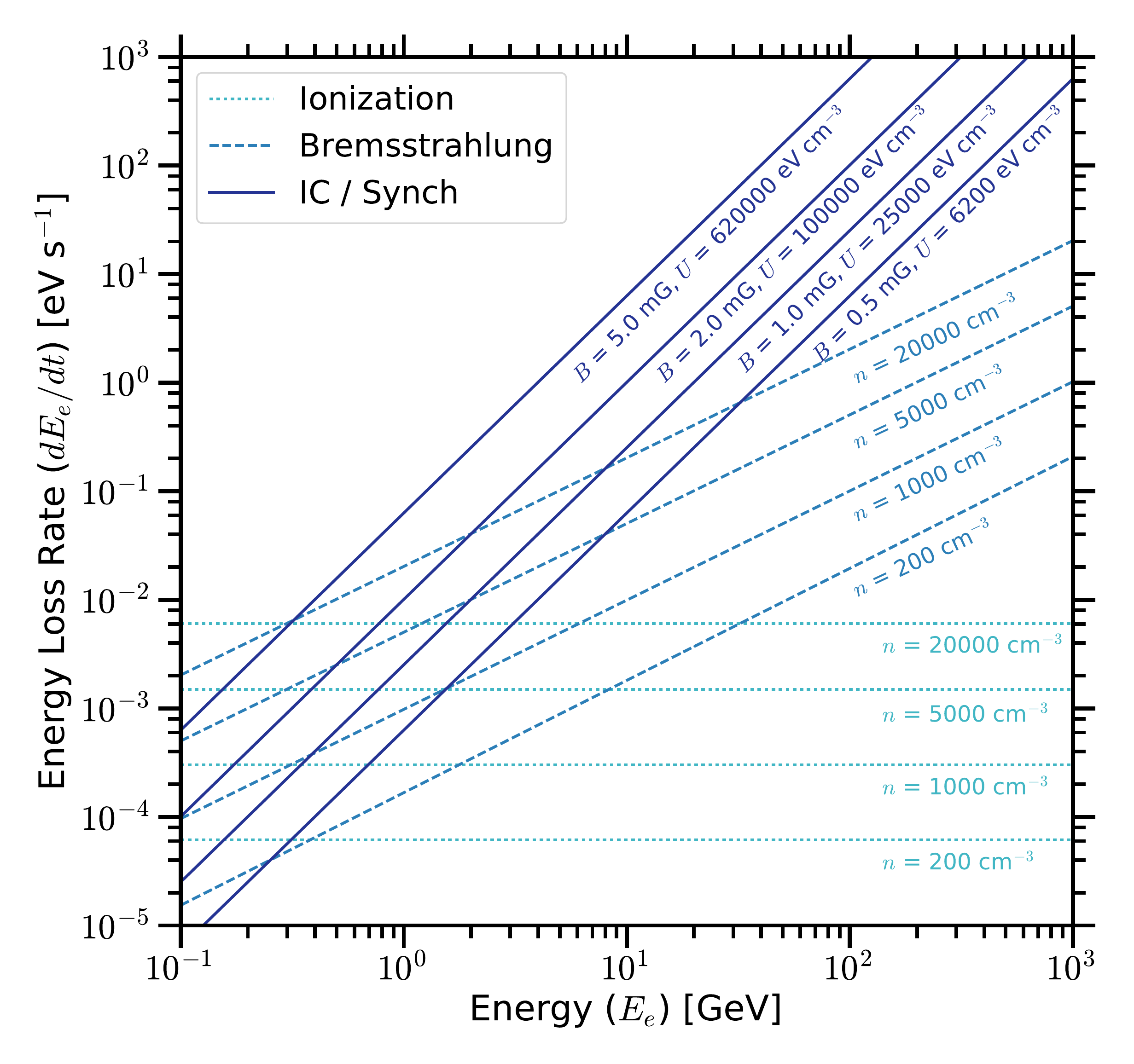}}
 \subfigure{
  \includegraphics[width=0.90\linewidth]{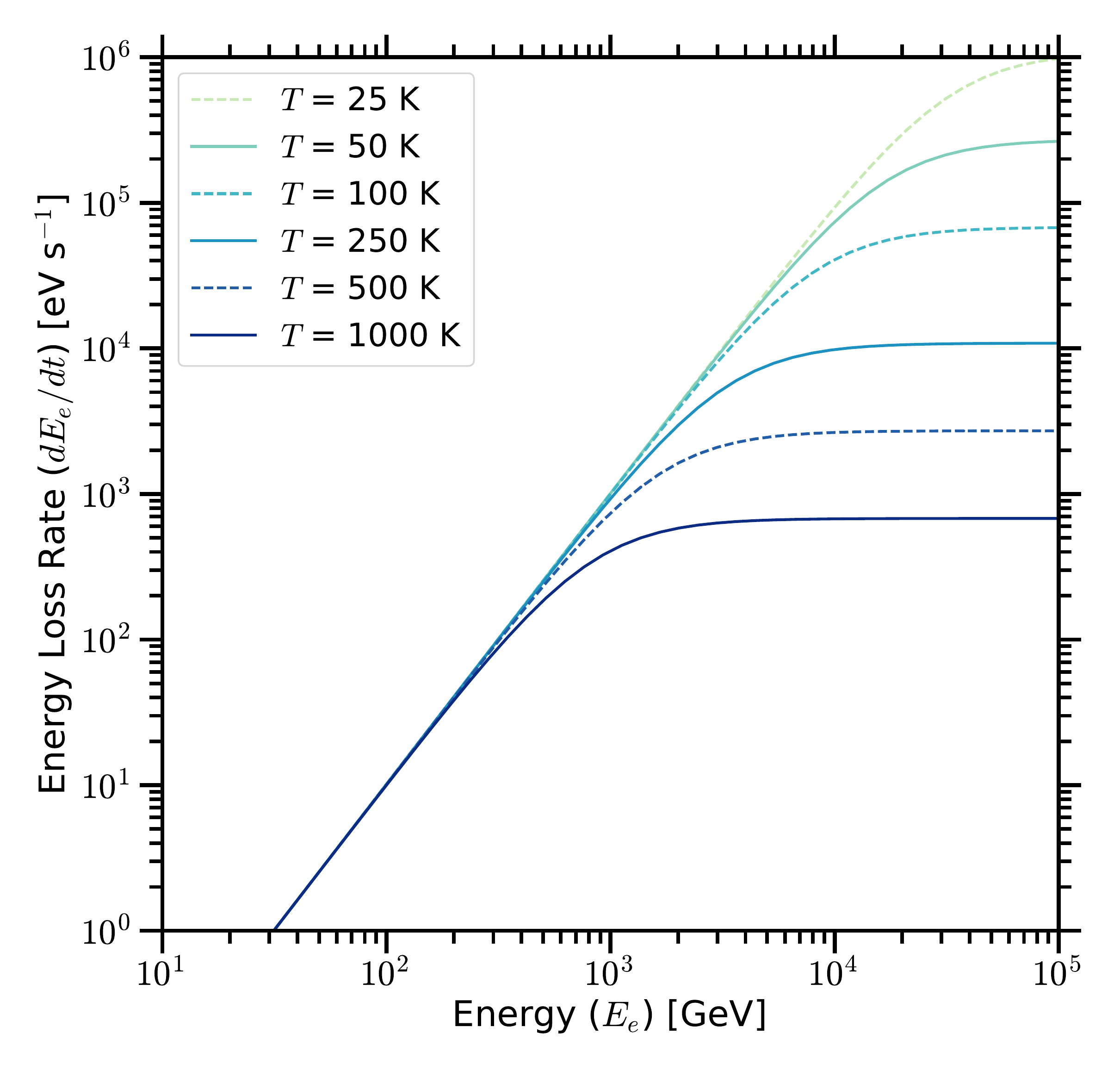}}
\caption{Top: Energy loss rates for cosmic ray leptons are shown, including losses due to ionization (dotted lines), bremsstrahlung (dashed lines), and synchrotron / inverse Compton (solid lines).  Bottom: Energy loss rates from inverse Compton are shown for varying temperature.  As the temperature increases, the energy at which the energy loss rate becomes energy independent decreases.}
\label{fig:lossrates}
\end{figure}

The cosmic ray confinement time is comprised of the energy-loss time-scale ($t_{\rm loss}(E_{e})$), the diffusion time-scale ($t_{\rm diff}(E_{e})$), and the energy-independent advective time-scale ($t_{\rm adv}$) such that
\begin{equation}
t_{e}(E_{e})^{-1} = t_{\rm loss}(E_{e})^{-1} + t_{\rm diff}(E_{e})^{-1} + t_{\rm adv}^{-1}.\label{eqn:timescale}
\end{equation}
Because of the extreme environment of Arp 220, the energy-loss time-scale is significantly shorter than either the diffusion or advective time-scales, see Fig.~\ref{fig:timescales}.  Further, an advective time-scale competitive with the energy-loss time-scale would only result in the need for a larger cosmic ray energy injection rate for the observed $\gamma$-ray spectrum, increasing the discrepancy between the injection rate derived from the $\gamma$-ray spectrum and the observed supernova rate.  Thus, to obtain a conservative lower limit on the cosmic ray injection rate, we will assume that the total lifetime for the cosmic ray electrons is simply
\begin{equation}
t_{e}(E_{e}) = t_{\rm loss} (E_{e}) \simeq E_{e} / (-dE / dt).\label{eqn:losstime}
\end{equation}

Based on the above equations, lowering the radio flux requires lowering either the synchrotron power ($P_{\nu}^{\rm Sy}$) or the lepton energy spectrum ($N_{e}(E_{e})$), see Fig.~\ref{fig:schematic}.  From equation~\ref{eqn:sypower}, we can see that the only means by which we can decrease the synchrotron power is by decreasing the magnetic field strength.  

The situation in regard to the energy spectrum for cosmic ray leptons is more complicated. To decrease the leptonic energy spectrum, we could decrease the energy injection spectrum.  However, the observed $\gamma$-ray spectrum provides stringent constraints on the cosmic ray spectral index ($\Gamma_{p}$) and a lower limit on the cosmic ray energy injection rate ($\varepsilon_{\rm CR}$).  Thus, lowering the population of secondary electrons and positrons is not possible within the confines of the observed $\gamma$-ray flux.  Hence, the primary avenue for lowering the leptonic energy spectrum is via the confinement time-scale, see Fig.~\ref{fig:schematic}.  As the energy-loss time-scale dominates the total confinement time-scale, see Fig.~\ref{fig:timescales}, we will concentrate our efforts on the energy loss rate for cosmic ray electrons.

\subsection{Energy Loss Rates} \label{sec:rates}

As in previous papers, we include energy losses via ionization, bremsstrahlung, inverse Compton, and synchrotron in our models.  The energy loss rates for these mechanisms are directly proportional to the gas density ($n$), the radiation field energy density ($U_{\rm rad}$), and the magnetic field strength ($B$) such that \citep{Ginzburg69,Schlick02,Dermer09,Schlick10}
\begin{align}
&-\left( \frac{dE_{e}}{dt} \right)_{\rm Ion} = \frac{9}{4} m_{e} c^{3} \sigma_{T} n_{\rm mol} \left[ 6.85 + \ln(\gamma_{e}) \right], \label{eqn:loss_ion} \\
&-\left( \frac{dE_{e}}{dt} \right)_{\rm Br}^{\rm ion} = \frac{3 \alpha}{2 \pi} c \sigma_{T} n_{\rm ion} Z (Z + 1) E_{e} \nonumber \\
& ~~~~~~~~~~~~~~~~~~~~ \times \left[ \ln(2\gamma_{e}) - \frac{1}{3} \right], \label{eqn:loss_bri} \\
&-\left( \frac{dE_{e}}{dt} \right)_{\rm Br}^{\rm mol} = \frac{3.9 \alpha}{8 \pi} c \sigma_{T} \phi_{\rm \ion{H}{i}}^{s-s} n_{\rm mol} E_{e}, \label{eqn:loss_brm} \\
&-\left( \frac{dE_{e}}{dt} \right)_{\rm IC} = \frac{4}{3} c \sigma_{T} U_{\rm rad} \frac{\gamma_{\rm K}^{2} \gamma_{e}^{2}}{\gamma_{\rm K}^{2} + \gamma_{e}^{2}}, \label{eqn:loss_ic} \\
&-\left( \frac{dE_{e}}{dt} \right)_{\rm Sy} = \frac{4}{3} c \sigma_{T} \frac{B^{2}}{8 \pi} \gamma_{e}^{2}, \label{eqn:loss_sy}
\end{align}
where $\phi_{\rm \ion{H}{i}}^{s-s} \approx 45$ is the scattering function for strongly shielded gas.  

The loss rates for ionization and bremsstrahlung both scale with the gas density (note that $n_{\rm mol} \gg n_{\rm ion}$) and because of the relative energy dependencies, bremsstrahlung losses dominate over ionization losses starting at $E_{e} \gtrsim 0.5$~GeV, see Fig.~\ref{fig:lossrates}.  Similarly, the loss rates for inverse Compton and synchrotron have the same energy dependence, and thus, inverse Compton losses dominate for $U_{\rm rad} > U_{B}$ and synchrotron losses dominate for $U_{\rm rad} < U_{B}$, where $U_{B} = B^{2} / 8 \pi$.

For `low' densities ($n \sim 200$~cm$^{-3}$) and `high' magnetic or radiation fields ($U > 10^{5}$~eV~cm$^{-3}$), synchrotron or inverse Compton losses dominate for all relevant energies (see Fig.~\ref{fig:lossrates}).  However, for `moderate' densities ($n \sim 1000$~cm$^{-3}$) and more `moderate' magnetic or radiation fields ($U > 1000$~eV~cm$^{-3}$), bremsstrahlung losses are dominate at low energies ($E_{e} < 1.0$~GeV) and synchrotron or inverse Compton losses dominate at higher energies, see Fig.~\ref{fig:lossrates}.

In equation~(\ref{eqn:loss_ic}), $\gamma_{\rm K}$ is the critical Klein-Nishina Lorentz factor which is given by $\gamma_{\rm K} = 3 \sqrt{5} m_{e} c^{2} / 8 \pi k T_{\rm rad}$ \citep{Schlick10}.  For $\gamma \ll \gamma_{\rm K}$, the inverse Compton cross section is well approximated by the Thompson cross section, while for $\gamma \geq \gamma_{\rm K}$ the full Klein-Nishina cross section must be used and the energy loss rate from inverse Compton becomes energy independent.  As shown in Fig.~\ref{fig:lossrates}, this change in the cross section becomes important for either high energies ($E_{e} \gtrsim 10$~TeV), high dust temperatures ($T_{\rm rad} \gtrsim 250$~K), or high optical depths.

\section{Observational Constraints} \label{sec:constraints}

As in previous work \citep{YoastHull15,YoastHull17a}, we model the eastern nucleus with a one-zone model and the western nucleus as two zones: an inner circumnuclear disc (CND) and an outer torus.  Here, we will discuss the relevant properties for both nuclei, and we have listed these properties in Table~\ref{tab:obs}.

Based on observations of the molecular media and high-resolution radio continuum measurements, we assume radii of $R = 85$~pc, $R = 65$~pc, and $R = 15$~pc for the eastern nucleus, the outer western nucleus, and the CND, respectively \citep{Barcos15,Scoville17}. We assume scale-heights of $h = 30$~pc in the eastern and outer western nuclei and $h = 10$~pc for the CND \citep{Scoville17,Sakamoto17}.

\subsection{Molecular Gas Masses} \label{sec:gasmass}

Millimeter wavelength observations provide probes of the H$_{2}$ column density ($N_{H_{2}}$) ranging from $10^{25}$ cm$^{-2}$ to greater than $10^{26}$ cm$^{-2}$ \citep{Wilson14,Scoville17}, assuming a Galactic CO-to-H$_{2}$ conversion factor of $X_{\rm CO, 20} = 2$.  Modeling and observations suggest that the conversion factor should be lower in ULIRGs, with potential values ranging from $X_{\rm CO, 20} \simeq 0.2 - 1.0$ \citep[][and references therein]{Bolatto13}.  We require that the molecular gas mass be less than the dynamical mass.  So, for a disc of radius $R$ and total height $H$, we calculate the molecular gas mass such that 
\begin{equation}
M_{\rm H_{2}} = \pi R H \times m_{\rm H_{2}} N_{\rm H_{2}} \frac{X_{\rm CO}^{\rm ULIRG}}{X_{\rm CO}^{\rm MW}}. \label{eqn:mass}
\end{equation}

For the eastern and outer western nuclei, we adopt moderate column densities of $10^{25}$~cm$^{-2}$, and for the CND, we assume a density of $10^{26}$~cm$^{-2}$ based on \citet{Scoville17}.  By choosing a moderate conversion factor of $X_{\rm CO, 20}^{\rm ULIRG} = 0.6$, we settle on molecular gas masses of $\sim 10^{9}$~M$_{\odot}$ for each of the nuclei, see Table~\ref{tab:obs}, which is consistent with other estimates of the gas mass \citep[e.g.,][]{Barcos15}.

\begin{table} 
\begin{minipage}{80mm}
\begin{center}
\caption{Fiducial Inputs}
\begin{tabular}{ccccc}
\hline
Parameters & East & West & CND & Ref.\\
\hline
Radius & \multirow{2}{*}{85} & \multirow{2}{*}{65} & \multirow{2}{*}{15} & \multirow{2}{*}{1,2} \\
$R$ [pc] & & & & \\[0.2cm]
Scale Height & \multirow{2}{*}{30} & \multirow{2}{*}{30} & \multirow{2}{*}{10} & \multirow{2}{*}{1} \\
$h$ [pc] & & & & \\[0.2cm]
Column Density & \multirow{2}{*}{$10^{25}$} & \multirow{2}{*}{$10^{25}$} & \multirow{2}{*}{$10^{26}$} & \multirow{2}{*}{1,3} \\
$N_{\rm H_{2}}$ [cm$^{-2}$] & & & & \\[0.2cm]
Molecular Gas Mass$^{a}$ & \multirow{2}{*}{10} & \multirow{2}{*}{5.9} & \multirow{2}{*}{5.9} & \multirow{2}{*}{4} \\
$M_{\rm H_{2}}$ [$10^{8} ~ \rm M_{\odot}$] & & & & \\[0.2cm]
Average ISM Density$^{a}$ & \multirow{2}{*}{1.1} & \multirow{2}{*}{1.2} & \multirow{2}{*}{65} & \\
$n_{\rm H_{2}}$ [$10^{4}$ cm$^{-3}$] & & & & \\[0.2cm]
FIR Luminosity & \multirow{2}{*}{3.0} & \multirow{2}{*}{3.0} & \multirow{2}{*}{6.0} & \multirow{2}{*}{1,3} \\
$L_{\rm IR}$ [$10^{11} ~ \rm L_{\odot}$] & & & & \\[0.2cm]
FIR Energy Density$^{a}$ & \multirow{2}{*}{2.8} & \multirow{2}{*}{4.2} & \multirow{2}{*}{150} & \\
$U_{\rm IR}$ [$10^{4}$ eV~cm$^{-3}$] & & & & \\[0.2cm]
Dust Temperature$^{a}$ & \multirow{2}{*}{50} & \multirow{2}{*}{55} & \multirow{2}{*}{135} & \multirow{2}{*}{} \\
$T_{\rm IR}$ [K] & & & & \\
\hline
\multicolumn{5}{l}{\textit{Notes.} $^{a}$Derived from above parameters; References:}\\
\multicolumn{5}{l}{(1) \cite{Scoville17}; (2) \cite{Barcos15};}\\
\multicolumn{5}{l}{(3) \cite{Wilson14}; (4) \cite{Bolatto13}}\\
\label{tab:obs}
\end{tabular}
\end{center}
\end{minipage}
\end{table}
%
%

\subsection{Radiation Fields} \label{sec:radfield}

Measurements of the infrared luminosity give us a means to calculate a lower limit on the radiation field energy density ($U_{\rm IR}$) such that
\begin{equation}
U_{\rm IR} = \frac{L_{\rm IR}}{4 \pi R^{2} \times c}. \label{eqn:uir}
\end{equation}
From the Stefan-Boltzmann law ($U_{\rm IR} = a T_{\rm IR}^{4}$), we can also calculate an effective dust temperature from the infrared luminosity such that
\begin{equation}
T_{\rm IR} = \left( \frac{L_{\rm IR}}{(2 \pi R^{2} + 2 \pi R H) \times c a} \right)^{1/4}, \label{eqn:temp}
\end{equation}
where $a$ is the radiation constant $a = 4 \sigma / c$.  

For the CND, estimates from millimeter wavelength observations put the infrared luminosity at $L_{\rm IR} \simeq 6 \times 10^{11}$~L$_{\odot}$ \citep{Wilson14,Scoville17}.  This corresponds to an effective dust temperature of 135~K with an energy density of $1.5 \times 10^{6}$~eV~cm$^{-3}$.  Adopting slightly less intense luminosities for the eastern nucleus and the outer western nucleus gives us effective dust temperatures of $\sim 50$~K and energy densities of $\sim 10^{4}$~eV~cm$^{-3}$, see Table~\ref{tab:obs}.

\subsection{Optical Depths} \label{sec:optdepth}

To lower the total radio emission while simultaneously preserving a high energy injection rate, one could increase the radiation field energy density (or the infrared optical depth), increase the average ISM density, or decrease the magnetic field strength.  From Zeeman splitting in OH masers in Arp 220, observations confirm that the magnetic field strengths in the Arp 220 nuclei are of mG strength \citep{McBride15}.  Magnetic fields of similar strengths are found even in extremely dense molecular gas clouds in our own Milky Way \citep[e.g.,][and references therein]{Crutcher12}.  Thus, we cannot decrease the magnetic field strength significantly from previous models \citep{YoastHull16}.  

Similarly, estimates of the dynamical masses for the nuclei are $\sim 2 \times 10^{9} ~ \rm M_{\odot}$ which provides a rough upper limit on the molecular gas mass \citep{Scoville17}.  Therefore, we cannot increase the average ISM density significantly above previous models \citep{YoastHull17a}.

Based on these observational constraints, we will be focusing on the optical depth and thus the radiation field energy density $U_{\rm rad}$.  For an optically thick disc, the radiation field energy density in the disc is higher than the emergent radiation field energy density, inferred from observations using equation~(\ref{eqn:uir}). The energy density and dust temperature in the disc are related to the emergent flux and radiation temperature such that
\begin{align}
U_{\text{rad}} &= \tau_{\rm IR} U_{\rm IR} \\
T_{\rm rad} &= \tau_{\rm IR}^{1/4} T_{\rm IR}.
\end{align}
The effective optical depth ($\tau_{\rm IR}$) used here is the optical depth of the integrated infrared spectrum, which peaks near 60~$\mu$m \citep{Gonzalez04}.  We will discuss the equivalent optical depth for millimeter wavelengths and compare with observational estimates in Section~\ref{sec:opacity}.

Using these relations, we find that for the optically thick case ($\tau_{\rm IR} \geq 1$)
\begin{align}
-\left( \frac{dE}{dt} \right)_{\rm IC} &\propto U_{\rm IR} \tau_{\rm IR}, \label{eqn:loss_tau} \\
q_{\rm IC}(E_{\gamma}) &\propto \frac{U_{\rm IR} \tau_{\rm IR}^{1/2}}{(kT_{\rm IR})^{2}}. \label{eqn:qtau}
\end{align}
Additionally, from equations~(\ref{eqn:qic2}) \& (\ref{eqn:qic_fin}), we know that the inverse Compton emissivity ($q_{\rm IC}$) depends on equation~(\ref{eqn:int}), $I(u \times \tau_{\rm IR}^{1/4} kT_{\rm IR},E_{\gamma})$, which depends on the electron loss time-scale and thus the electron energy loss rates, see Appendix~\ref{sec:appb} for details.

\section{Results} \label{sec:results}

\subsection{Model Setup} \label{sec:setup}

The goal of this study is to find a solution or set of solutions in parameter space via $\chi^{2}$ tests that simultaneously fit the $\gamma$-ray and radio spectra.  For the $\gamma$-ray spectra, the most critical parameters are the cosmic ray spectral index and the cosmic ray energy injection rate.  Because the $\gamma$-ray observations provide no information about regarding spatial distribution, we test three different scenarios for where and how energy is injected into cosmic rays.

Previously, we began with injection rates of $0.7 \times 10^{50}$~erg~yr$^{-1}$, $0.7 \times 10^{50}$~erg~yr$^{-1}$, and $1.3 \times 10^{51}$~erg~yr$^{-1}$ for the eastern nucleus, western nucleus, and CND, respectively \citep{YoastHull15,Varenius16}.  Thus, in Scenario A, we begin by assuming similar initial injection rates of $0.5 \times 10^{50}$~erg~yr$^{-1}$, $0.5 \times 10^{50}$~erg~yr$^{-1}$, and $10^{50}$~erg~yr$^{-1}$ and increase the injection rates in equal proportion up to a total injection rate of $4 \times 10^{51}$~erg~yr$^{-1}$, see Table~\ref{tab:input}.  This model includes primary protons along with primary and secondary electrons, with the assumption that the primary electron-to-proton (number) ratio is $\xi = 0.02$ \citep{YoastHull13}.

\begin{table} 
\begin{minipage}{80mm}
\begin{center}
\caption{Cosmic Ray Energy Injection Rates}
\begin{tabular}{cccc}
\hline
& East & West & CND\\
\hline
\multicolumn{4}{c}{Scenario A (Protons \& Electrons)}\\
\hline
Range & $0.5 - 10$ & $0.5 - 10$ & $1.0 - 20$\\
Step Size & 0.5 & 0.5 & 1.0\\
\hline
\multicolumn{4}{c}{Scenario B (Protons Only)}\\
\hline
Range & 1.0 & 1.0 & $2.0 - 40$\\
Step Size & 0 & 0 & 2.0\\
\hline
\multicolumn{4}{c}{Scenario C (Electrons Only)}\\
\hline
Range & 1.0 & 1.0 & $10^{1} - 10^{5}$\\
Step Size & 0 & 0 & $10^{0.2}$\\
\hline
\multicolumn{4}{l}{\textit{Notes:} In all cases, the units for the injection rates}\\
\multicolumn{4}{l}{are $10^{50}$~erg~yr$^{-1}$.} 
\label{tab:input}
\end{tabular}
\end{center}
\end{minipage}
\end{table}

In Scenario B, we fix the injection rates in the eastern and outer western nuclei to $10^{50}$~erg~yr$^{-1}$ in each region to be consistent with the observed supernova rates, see \citet{Varenius16}.  In the CND, we test a range of injection rates from $2 \times 10^{50}$ to $4 \times 10^{51}$~erg~yr$^{-1}$, see Table~\ref{tab:input}, and inject only primary protons, including secondary electrons and positrons but not primary electrons.  In Scenario C, we similarly fix the injection rates in the eastern and western nuclei and test a range of rates from $10^{51} - 10^{55}$~erg~yr$^{-1}$ in steps of $10^{0.2}$~yr$^{-1}$ in the CND, see Table~\ref{tab:input}.  For Scenario C, we only include primary electrons and no primary protons or secondary electrons and positrons.

In each scenario, we test the models against both the $\gamma$-ray and radio spectra while varying the injection spectral index (2.3, 2.4, 2.5), magnetic field strength ($0.1 - 100$ mG), and optical depth ($1 - 10^{4}$).  For the radio spectra, we also vary the absorption fraction ($0.2 - 1.0$) and the ionized gas density ($50 - 800$ cm$^{-3}$), both of which affect thermal radio emission.  For our $\chi^{2}$ tests against the $\gamma$-ray data, we assume values for the spectral index and optical depth are equal across each of the regions (East, West, CND) in order to reduce the computation time, and we assume that the magnetic field strength in the CND is $\sim 7$ times larger than in the East and the West based on the difference in ISM densities.

Regarding radio data, there exist separate spectra for the eastern and western nuclei.  Thus, in Scenario A, we assume separate values for spectral index, optical depth, magnetic field, ionized gas density, and absorption fraction in the eastern nucleus versus the western nucleus, and we also assume that the spectral index, optical depth, and absorption fraction are constant across the outer western nucleus and the inner CND.  We also assume that the ionized gas density increases by a factor of $\sim 50$ from the outer western nucleus to the inner CND.  

In Scenarios B \& C, we also assume that the bulk of the $\gamma$-ray emission and the minority of the radio emission comes from the CND.  Thus, for the CND, we require that the radio emission be less than 1\% of the total emission from the western nucleus.  As such, we can test separate spectral indices, magnetic field strengths, and optical depths for the West versus the CND.

\begin{figure}
 \subfigure{
  \includegraphics[width=0.85\linewidth]{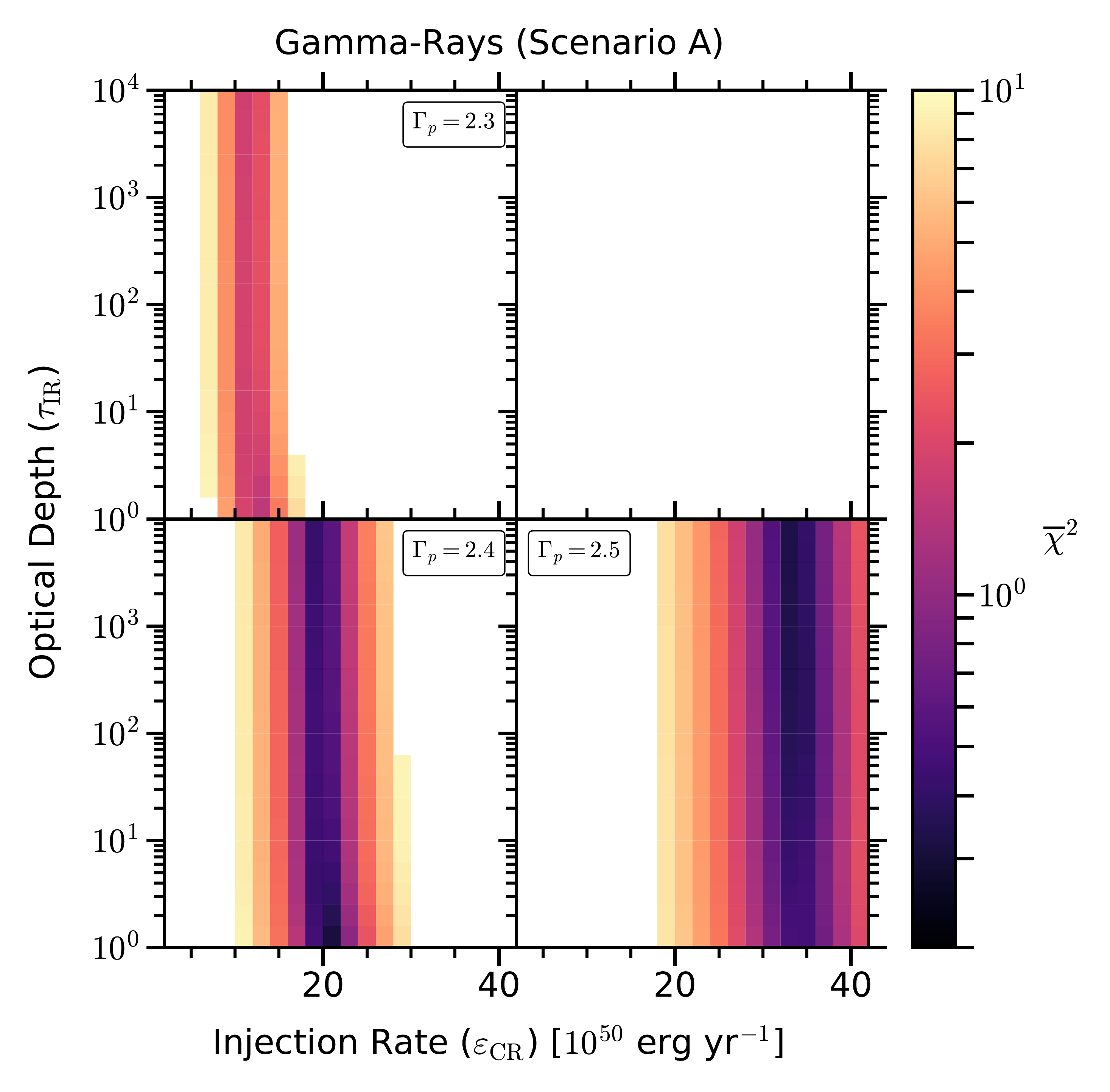}}
 \subfigure{
  \includegraphics[width=0.85\linewidth]{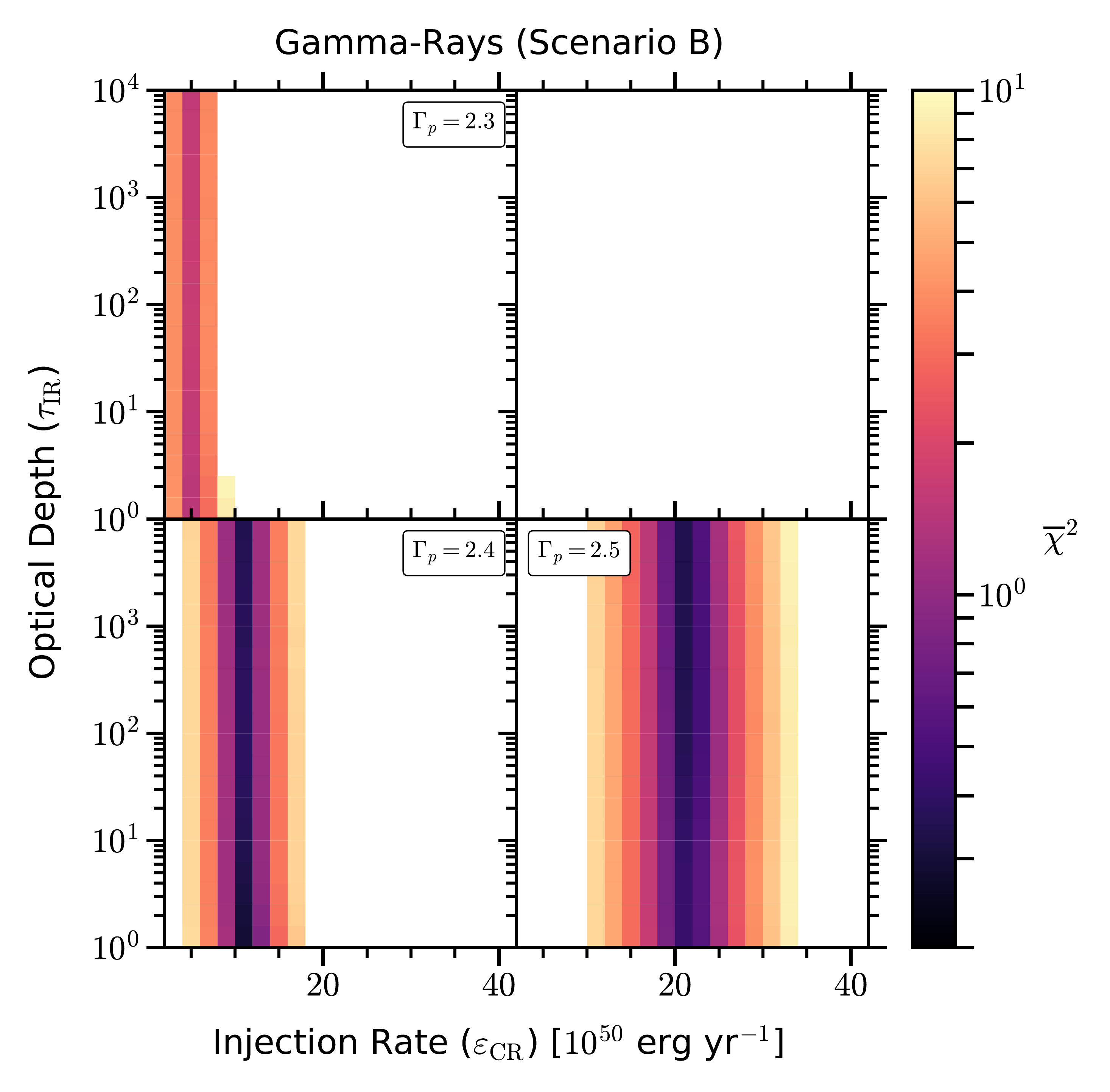}}
\caption{Chi-squared contours for Scenarios A and B for $\gamma$-ray data, injection rate ($\varepsilon_{CR}$) versus optical depth ($\tau_{\rm IR}$)  The $\gamma$-ray emission in Scenarios A and B is dominated by hadrons.  Thus, the results are not dependent on magnetic field strength or optical depth.}
\label{fig:gchi}
\end{figure}
%
%

\subsection{Gamma-Ray Spectra} \label{sec:gammarays}

\begin{figure}
 \subfigure{
  \includegraphics[width=0.85\linewidth]{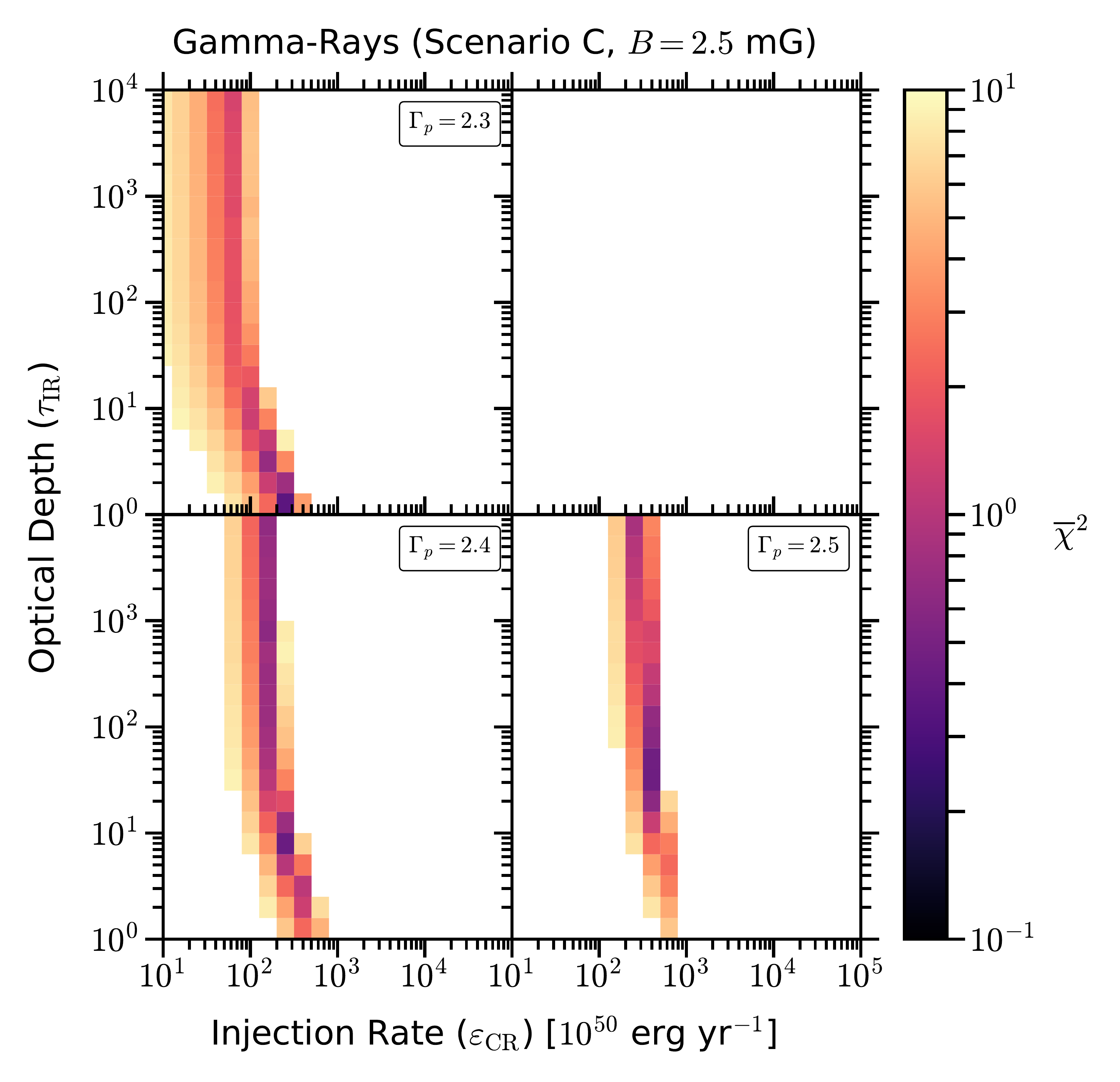}}
 \subfigure{
  \includegraphics[width=0.85\linewidth]{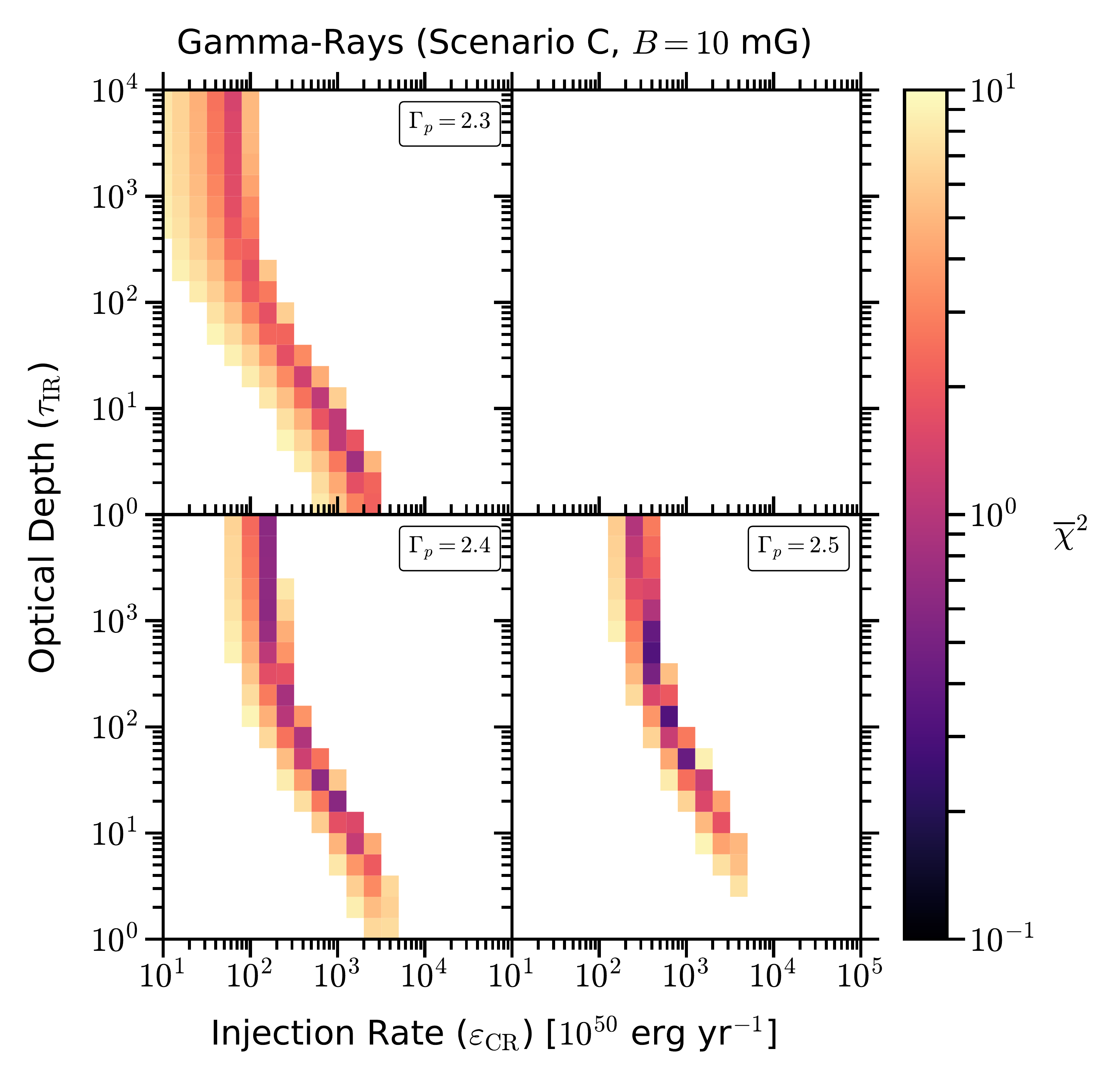}}
 \subfigure{
  \includegraphics[width=0.85\linewidth]{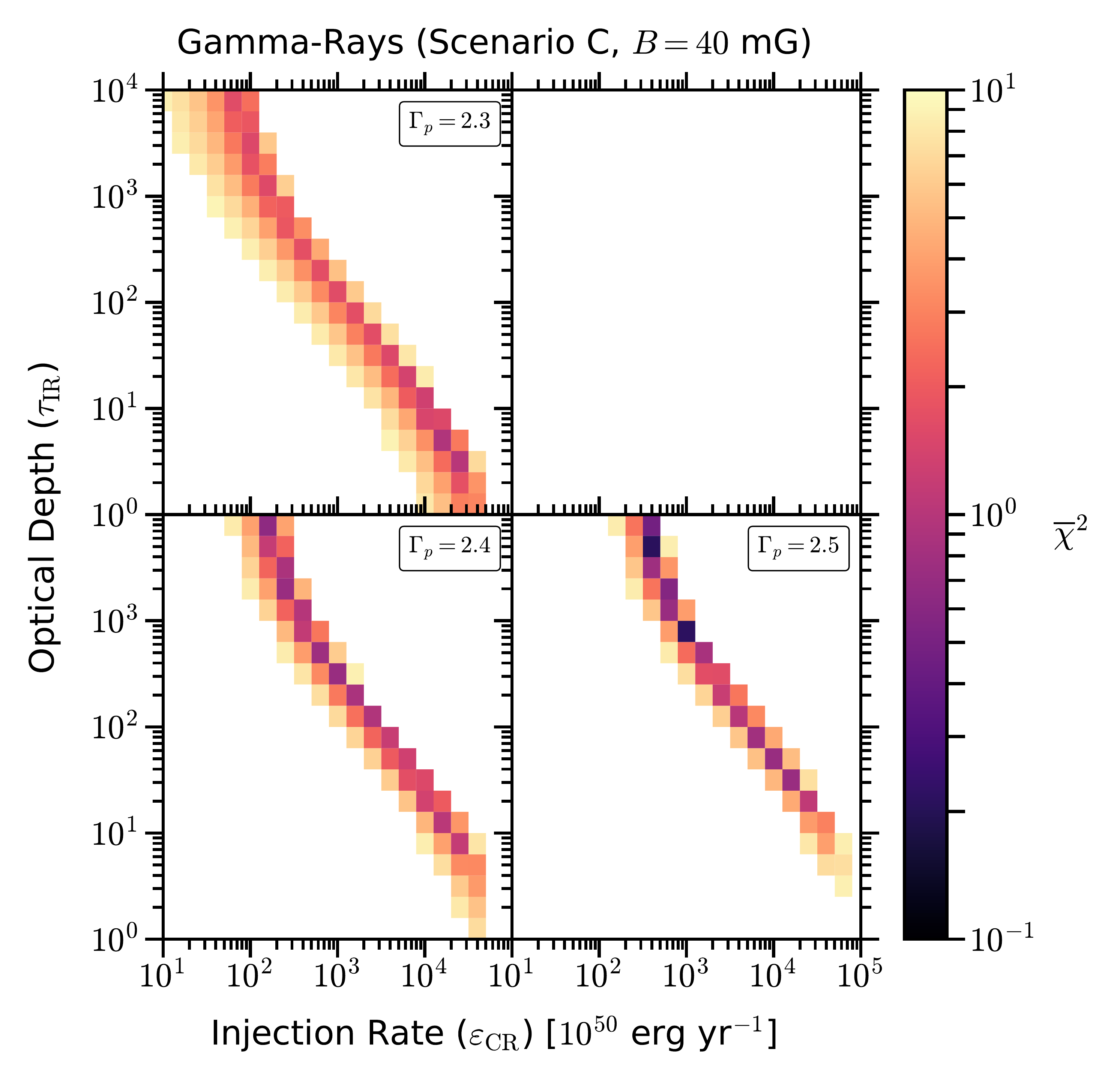}}
\caption{Chi-squared contours for Scenario C for $\gamma$-ray data, injection rate ($\varepsilon_{CR}$) versus optical depth ($\tau_{\rm IR}$).  The $\gamma$-ray emission in Scenario C is dominated by leptons.  Thus, as magnetic field strength increases, the dependency of the energy injection rate on optical depth increases.}
\label{fig:gchi2}
\end{figure}

The best-fitting results from the $\chi^{2}$ tests, performed against the $\gamma$-ray data from \citet{Peng16}, require energy injection rates of $2 \times 10^{51}$~erg~yr$^{-1}$ and $10^{51}$~erg~yr$^{-1}$ for Scenarios A and B, respectively, see Fig.~\ref{fig:gchi}.  These results for Scenarios A and B are independent of optical depth (see Fig.~\ref{fig:gchi}), as hadronic $\gamma$-rays dominate the emission in these models.  Additionally, these models are best fit with spectral indices of $\Gamma_{p} \simeq 2.4$, closely matching the spectral index ($\Gamma_{\gamma} = 2.35$) derived from fitting a pure power-law to the $\gamma$-ray spectrum \citep{Peng16}.  

For Scenario C, we assume injection of cosmic ray leptons, and thus the energy injection rate required to fit the $\gamma$-ray data changes with both optical depth and magnetic field strength, see Fig.~\ref{fig:gchi2}.  Overall, preferred models have energy injection rates of $\leq 10^{52}$~erg~yr$^{-1}$, and the best-fitting models require both high optical depths and high magnetic field strengths.  

Examining Fig.~\ref{fig:gchi2}, we can see that at low magnetic field strengths, the energy injection rate is nearly independent of optical depth.  However, as magnetic field strength increases, the relationship between optical depth and injection rate nears a pure power-law.  Further, these models show a clear requirement for an optical depth significantly greater than unity, see Fig.~\ref{fig:gchi2}.

\subsection{Radio Spectra} \label{sec:radio}

\begin{figure*}
 \subfigure{
  \includegraphics[width=0.425\linewidth]{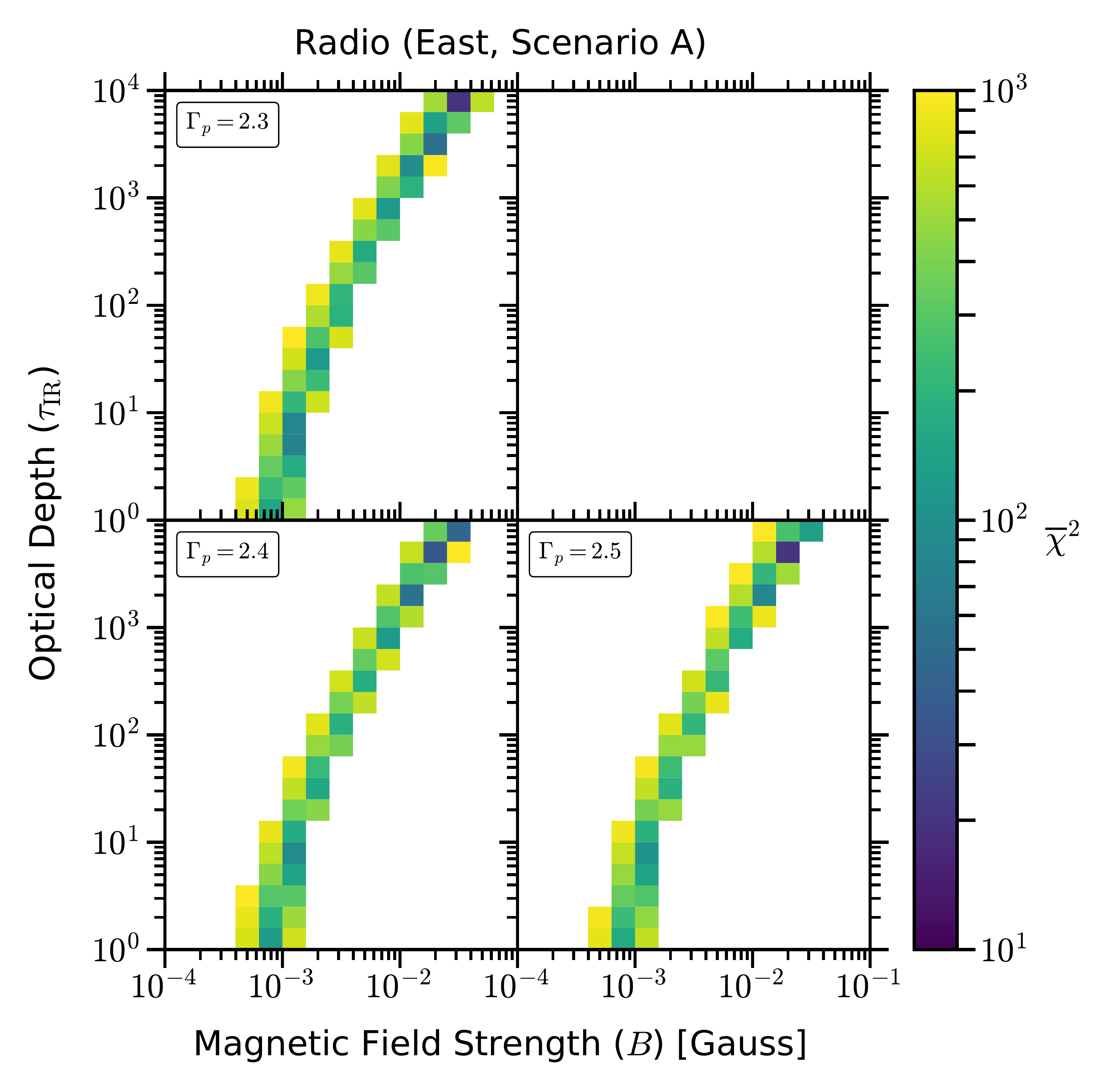}}
 \subfigure{
  \includegraphics[width=0.425\linewidth]{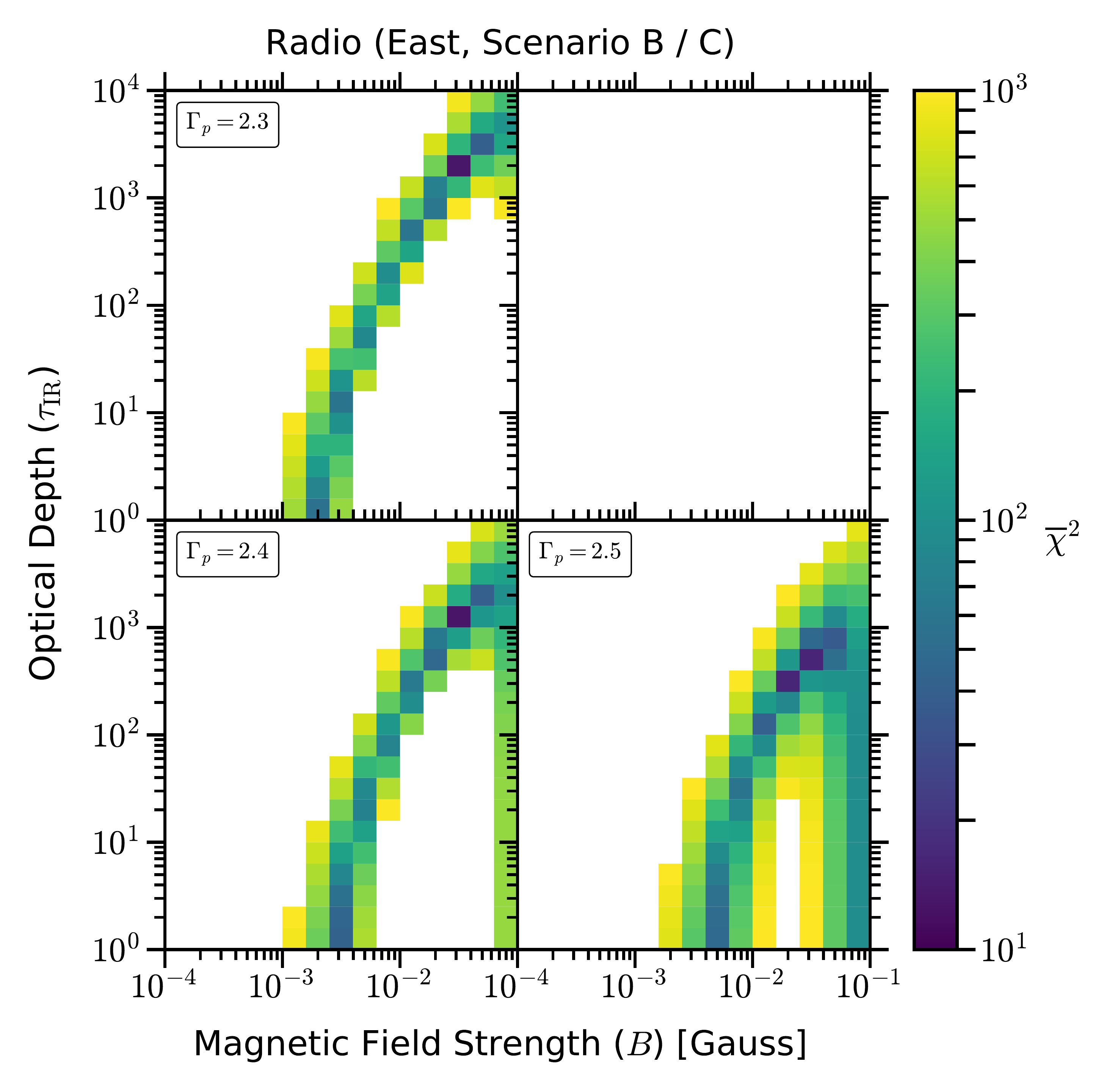}}
 \subfigure{
  \includegraphics[width=0.425\linewidth]{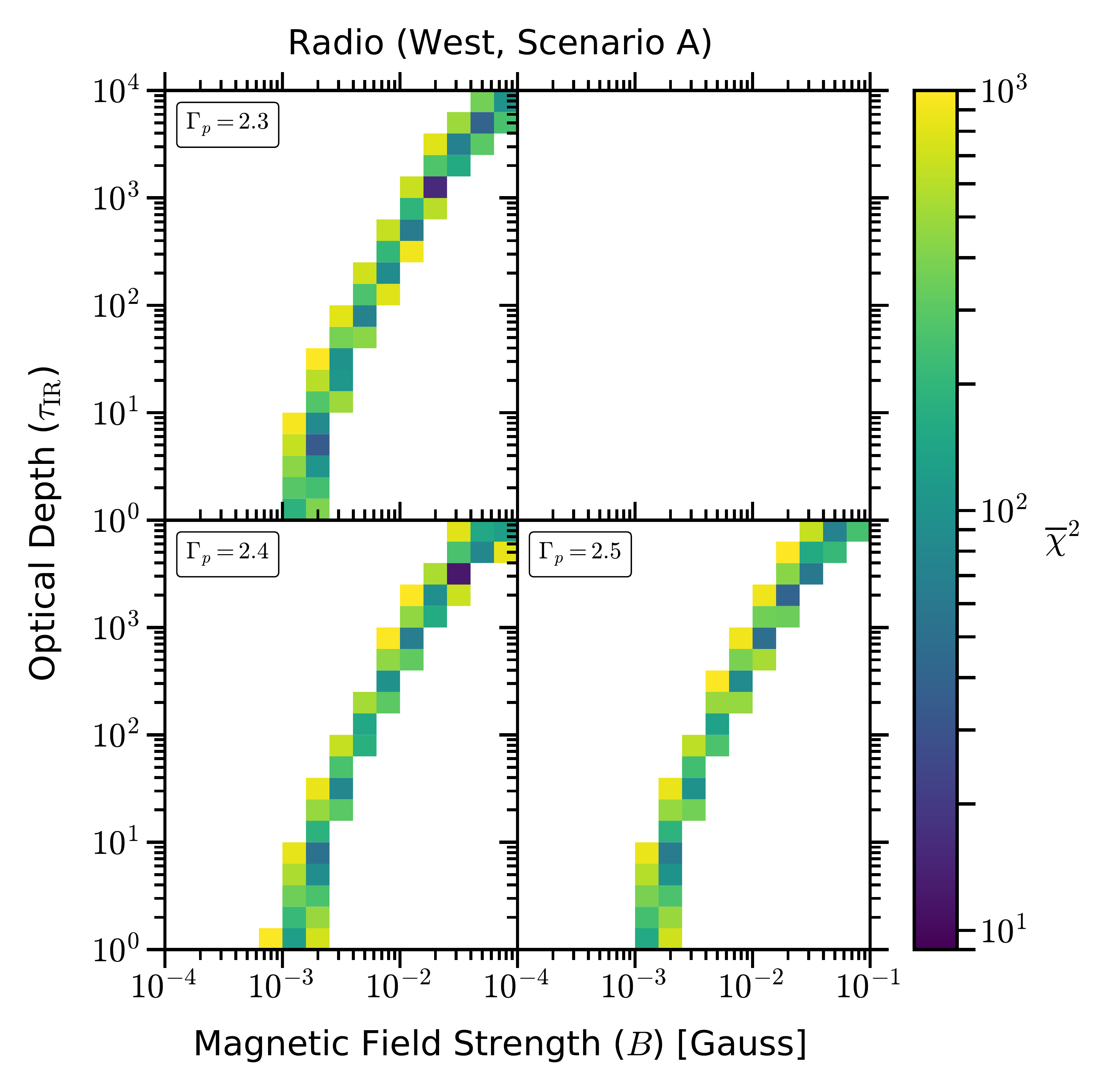}}
 \subfigure{
  \includegraphics[width=0.425\linewidth]{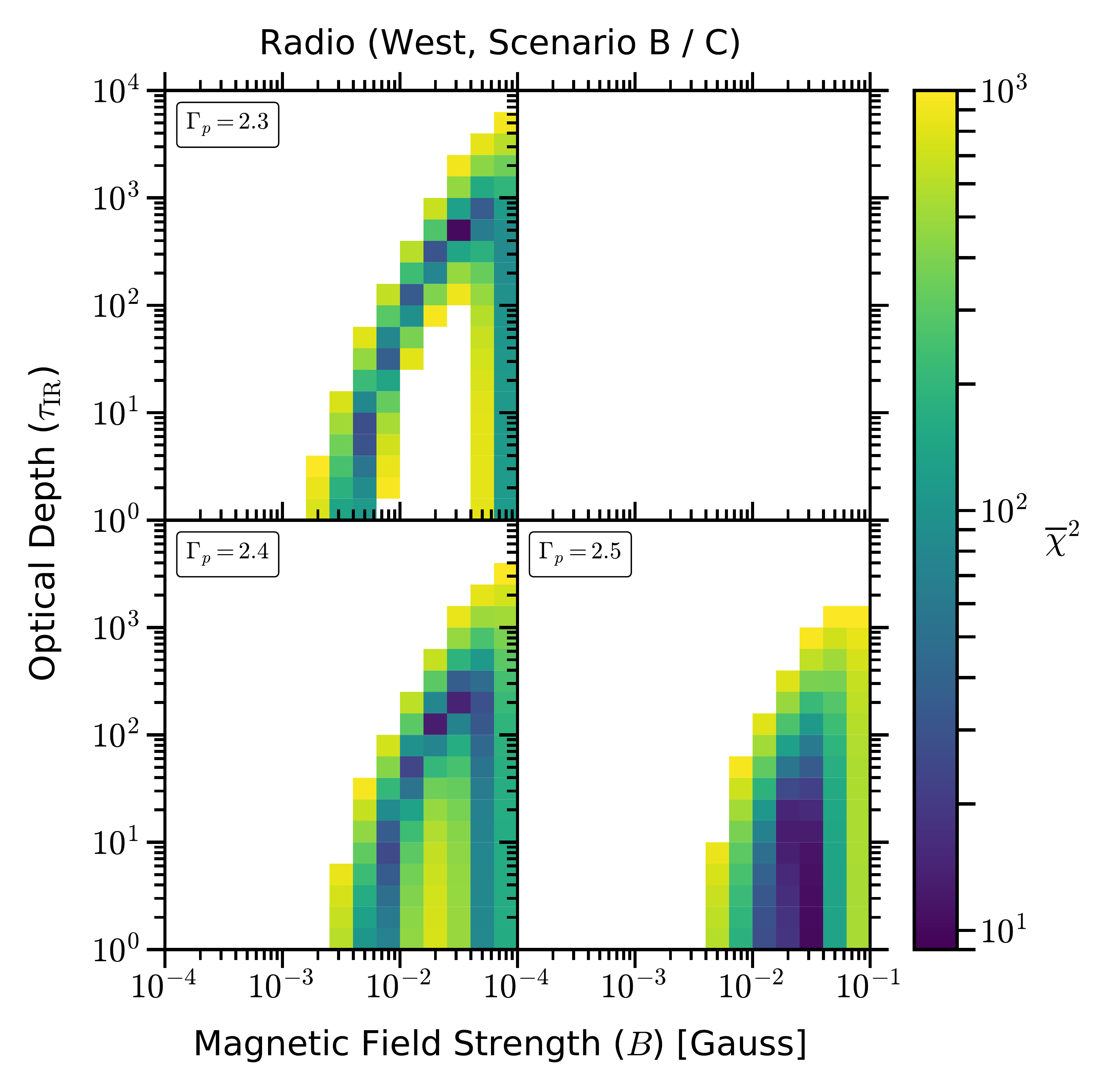}}
 \subfigure{
  \includegraphics[width=0.425\linewidth]{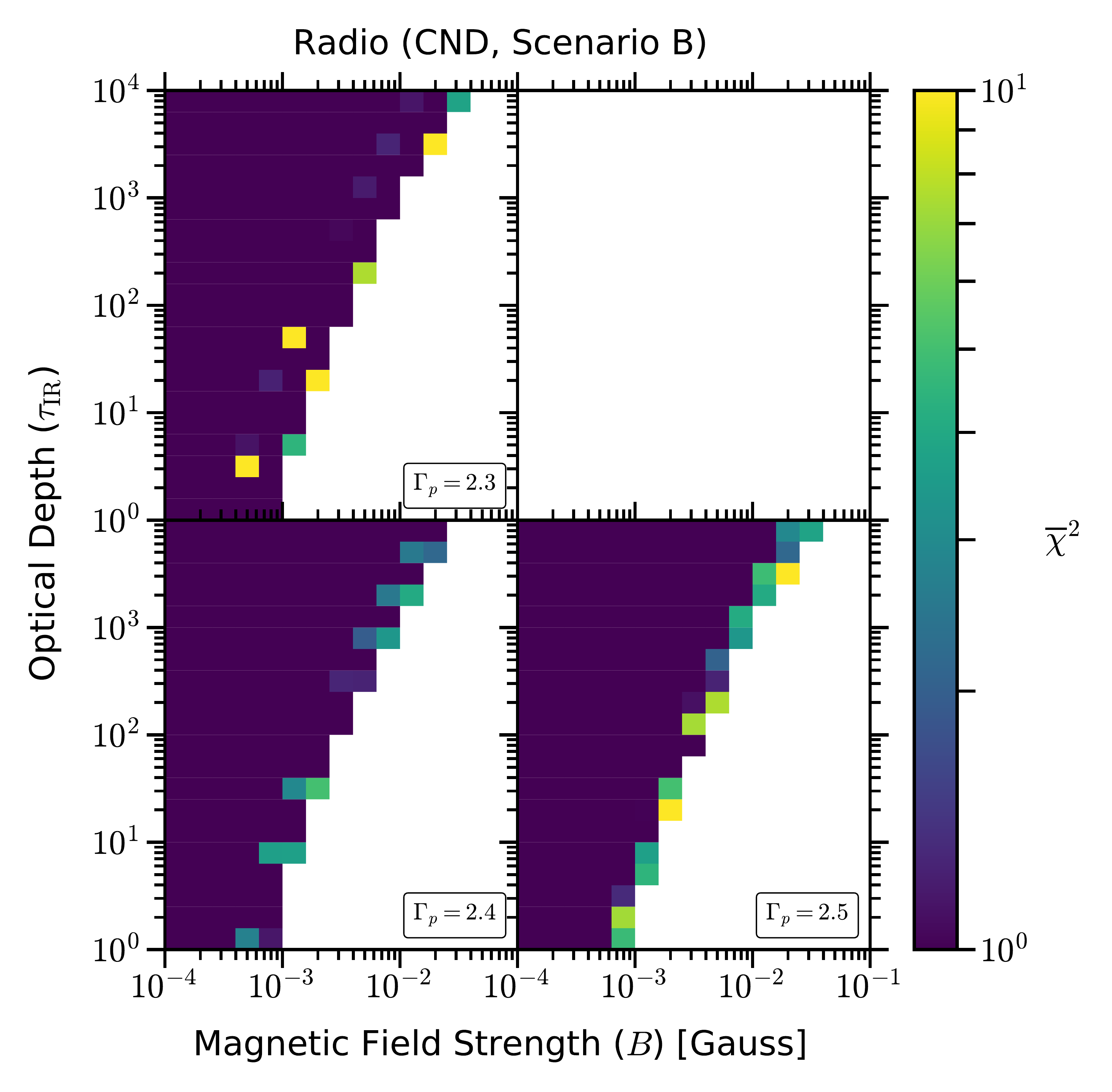}}
 \subfigure{
  \includegraphics[width=0.425\linewidth]{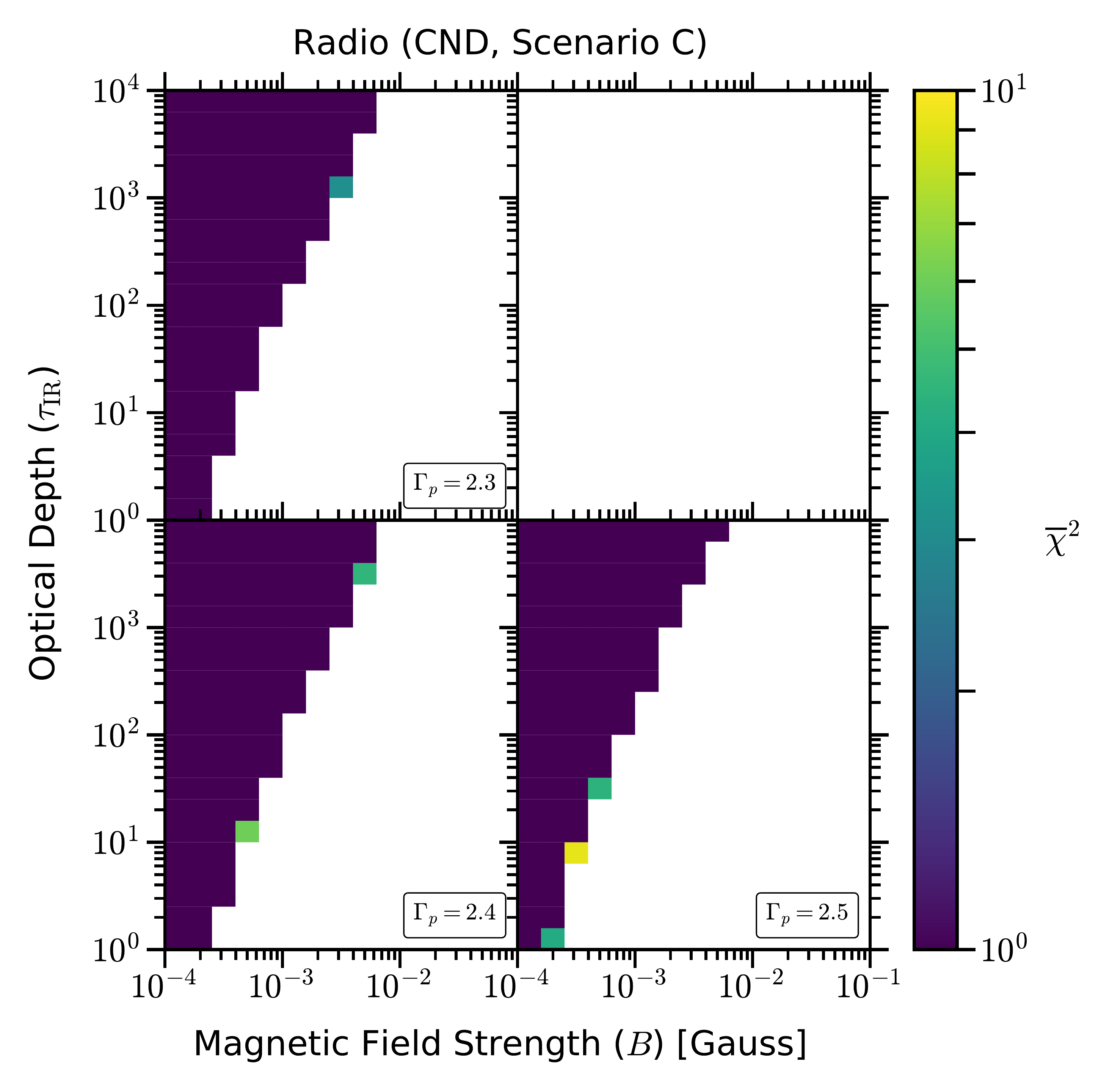}}
\caption{Chi-squared contours for radio data, magnetic field strength ($B$) versus optical depth ($\tau_{\rm IR}$).}
\label{fig:rchi}
\end{figure*}

For the radio spectra, we have collected data from \citet{Downes98,Rodriguez05,Barcos15}, covering a frequency range from 1.7 to 43 GHz. Here, we limit our analysis to models with only the best-fitting energy injection rates from the $\gamma$-ray analysis.  For Scenario A, our best-fitting models for both the eastern and western nuclei are restricted to magnetic field strengths greater than $10$ mG and optical depths greater than $10^{3}$, see Fig.~\ref{fig:rchi}.  For Scenarios B and C, where we have assumed constant injection rates of $10^{50}$~erg~yr$^{-1}$ for the East and West, we find solutions with magnetic field strengths as low as a few mG and optical depths ranging from unity to $\sim 10^{3}$.

In the case of the CND, we consider any model producing radio emission that is less than 1\% of the total radio emission from the western nucleus to be acceptably negligible in regards to the total radio emission.  These models are marked as having a $\chi^{2}$ value of 1 (corresponding to dark purple) in Fig.~\ref{fig:rchi}.  If we allow for magnetic field strengths below 1 mG, then the minimum requirement for the optical depth for a joint solution is simply unity.  If we require that the magnetic field strength be greater than or equal to 1 mG, then the optical depth must be larger than 10 for Scenario B and larger than 100 for Scenario C, see Fig.~\ref{fig:rchi}.  

In all cases, it is clear that as magnetic field strength increases, optical depth must also increase, see Fig.~\ref{fig:rchi}.  This is expected because as magnetic field strength increases, the total synchrotron emissivity increases.  Thus, to keep the radio emission small, the optical depth must also increase.  Above, we also found that for Scenario C, as the energy injection rate increases and assuming high ($\geq 10$ mG) magnetic field strengths, the optical depth must decrease, see Fig.~\ref{fig:gchi2}.  For the $\gamma$-rays, the $\gamma$-ray flux increases as the energy injection rate increases.  Thus, to avoid over-predicting the $\gamma$-ray flux, the optical depth must decrease as the energy injection rate increases. Regardless, there is clearly a minimum optical depth necessary to provide a joint solution to the radio and $\gamma$-ray spectra.

\section{Discussion} \label{sec:discussion}

\subsection{Observed Optical Depths} \label{sec:opacity}

As stated above, the best-fitting optical depths derived from our $\chi^{2}$ analysis are effective optical depths over the whole of the infrared range.  Thus, a full and complete comparison between our $\tau_{\rm IR}$ and optical depth values derived at specific wavelengths from observational measurements is complex.  However, a rough comparison is simple enough if we assume that our effective optical depth is approximately representative of the optical depth at the peak of the infrared spectrum, $\sim 60$ $\mu$m in Arp 220 \citep{Gonzalez04,Rangwala11}.  Assuming optical depth depends on frequency to the 1.6 power in ULIRGs \citep{Casey12}, we have
\begin{equation} \label{taumm}
\tau_{\rm IR} \approx \tau_{\nu} \left( \frac{\lambda(\nu)}{60~\mu m} \right)^{1.6}.
\end{equation}

In \citet{Wilson14}, the authors calculated optical depths of 1.7 and 5.3 for the eastern and western nuclei, respectively.  The optical depths were derived from a combination of submillimeter observations with the Atacama Large Millimeter/submillimeter Array (ALMA) at 691 GHz (434 $\mu$m) and at 345 GHz (869 $\mu$m).  Converting these optical depths to optical depths in the infrared range, we get $\tau_{60} = 40.3$ and $\tau_{60} = 125.7$ for the eastern and western nuclei, respectively.  These values would suggest that Scenarios B and C are more likely than Scenario A, as optical depths larger than $10^{3}$ were necessary for joint solutions between the radio and $\gamma$-ray data in Scenario A.

\subsection{Energy Densities} \label{sec:energydens}

From our results, it is clear that there is significantly higher energy density in infrared radiation in the nuclei of Arp 220 than in the free-streaming emergent (observed) radiation field.  This is consistent with submillimeter and millimeter wavelength estimates of the optical depth at those wavelengths \citep[e.g.,][]{Wilson14,Scoville17}.  To evaluate the how the energy density in the trapped radiation field compares with the other components of the ISM, we evaluate the energy densities for turbulence, magnetic fields, radiation fields, and cosmic rays below.

The turbulent energy density is given as $U_{\rm turb} = \rho \sigma_{\rm turb}^{2} / 2$, where $\rho$ is the ISM gas density ($\rho_{\rm H_{2}} = m_{\rm H_{2}} n_{\rm H_{2}}$).  Measurements by \citet{Scoville17} show the turbulent velocities to be 120 km~s$^{-1}$ and 250 km~s$^{-1}$ in the eastern and western nuclei, respectively.  Combined with gas densities of $\sim 10^{4}$ and $6.5 \times 10^{5}$ cm$^{-3}$, these velocities give us turbulent energy densities of $1.7 \times 10^{6}$ and $4.2 \times 10^{8}$ eV~cm$^{-3}$.

For magnetic field strengths of $\sim 4 - 20$ mG, the corresponding energy densities ($U_{B} = B^{2} / 8 \pi$) range from $4 \times 10^{5}$ eV~cm$^{-3}$ to $10^{7}$ eV~cm$^{-3}$.  Thus, the magnetic field energy density lags slightly behind the turbulent energy density in the eastern and outer western nuclei.  However, in the CND, the magnetic field energy density clearly lags behind the turbulent energy density by more than an order of magnitude.

For the radiation field, the total energy density is given by $U_{\rm rad} = \tau_{\rm IR} U_{\rm IR}$.  Thus, for the East and the West, the total radiation field energy densities range from $\sim 10^{4} - 10^{7}$ eV~cm$^{-3}$ based on the results of our $\chi^{2}$ analysis in Scenarios B and C.  In the CND, the total radiation field energy density must be greater than $\sim 10^{7}$ eV~cm$^{-3}$ and greater than $\sim 10^{8}$ eV~cm$^{-3}$ in Scenarios B and C, respectively.  

If we take our infrared optical depths approximated from the \citet{Wilson14} measurements, the corresponding energy densities are $1.1 \times 10^{6}$ eV~cm$^{-3}$ for the eastern / outer western nucleus versus $1.9 \times 10^{8}$ eV~cm$^{-3}$ for the CND.  These specific values are within the range of energy densities from our acceptable models and they agree well with the turbulent energy densities.

Finally, for the cosmic rays, the energy density involves an integral over the energy spectrum such that $U_{\rm CR} = \int E N(E) dE$.  For Scenarios B and C, the cosmic ray energy densities for the East and West are $1.8 \times 10^{3}$ and $3 \times 10^{3}$ eV~cm$^{-3}$.  Because Scenarios B and C have low assumed energy injection rates, along with high energy loss rates for cosmic rays, the corresponding energy densities are low and consistent with our previous study in \citet{YoastHull15}.  For the CND, in Scenarios B and C, the total cosmic ray energy densities are $\sim 4 \times 10^{4}$ eV~cm$^{-3}$.  Thus, even with the additional cosmic ray population inferred from the $\gamma$-ray flux, the cosmic ray energy density is well below the other energy densities.

\begin{figure}
 \subfigure{
  \includegraphics[width=0.90\linewidth]{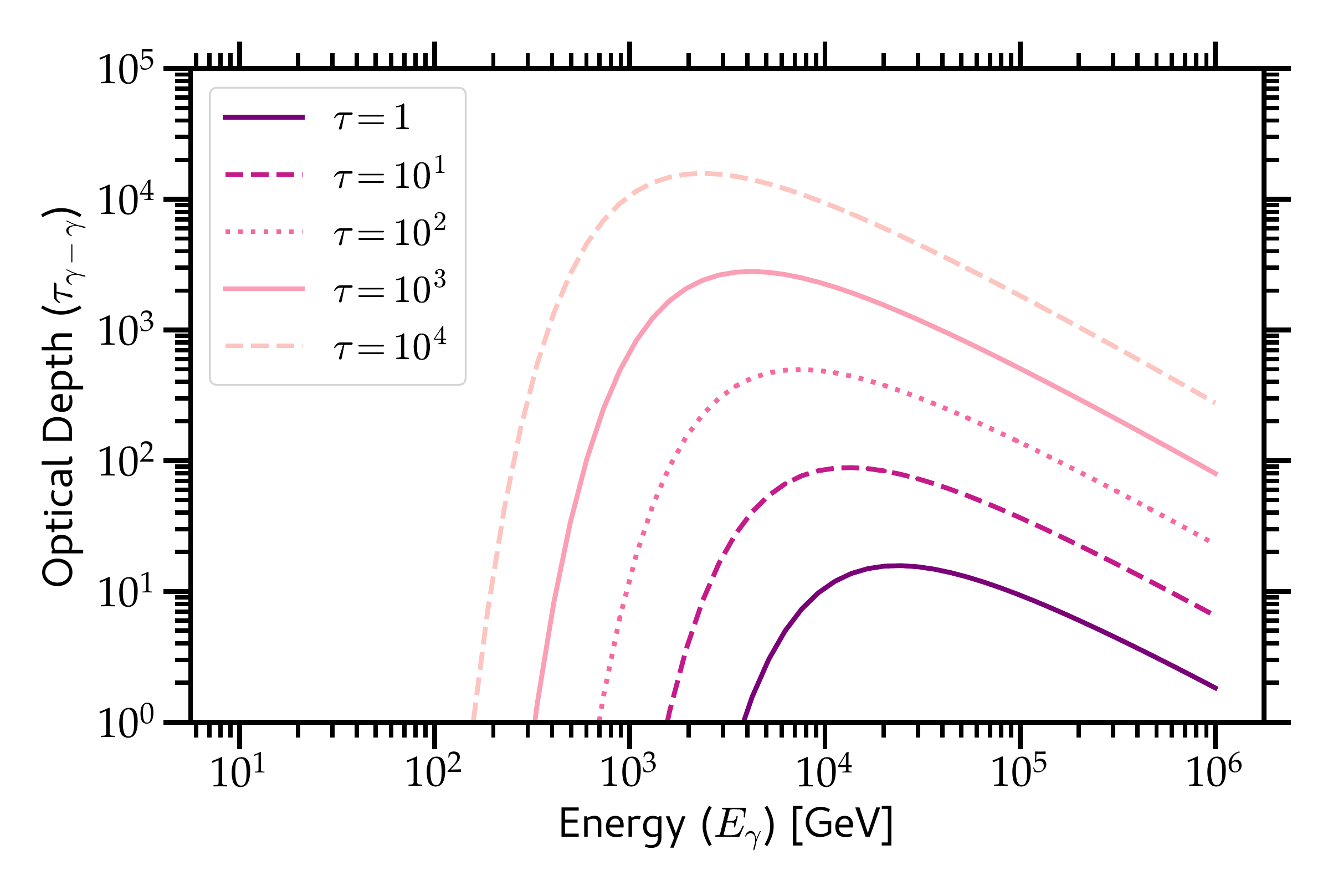}}
\caption{Optical depths for photon-photon interactions are shown for differing values of the infrared optical depth.  As the infrared optical depth increases, the photon-photon optical depth also increases and the peak shifts down in energy.  However, even at for very high infrared optical depth values, the photon-photon optical depth is not greater than unity below 100 GeV.}
\label{fig:pairp}
\end{figure}
%
%

\subsection{Further Interactions with Infrared Photons} \label{sec:int}

Having established that the infrared optical depth in the nuclei of Arp 220 is greater than unity, we must assess how this increase in the radiation energy density affects other interaction rates. Specifically, we examine how the optical depth for photon-photon interactions and the source function for photopion production increase with increasing infrared optical depth.

\subsubsection{Photon-Photon Interactions} \label{sec:ggint}

From previous studies, we know that even considering only the emergent infrared energy density, the radiation fields in Arp 220 are large enough for photon-photon ($\gamma - \gamma$) interactions to create significant absorption in the TeV energy $\gamma$-ray spectrum \citep{Torres04,Lacki13b,YoastHull15}.  The optical depth for $\gamma - \gamma$ interactions depends on the integral over the photon number density over the square of the photon energy and is thus directly proportional to the radiation field energy density and inversely proportional to the temperature to the third power, see Appendix~\ref{sec:appc} for details.  Hence, $\tau_{\gamma - \gamma}$ depends on the effective infrared optical depth such that $\tau_{\gamma - \gamma} \propto \tau_{\rm IR}^{1/4}$.

Calculating the photon-photon optical depth for various infrared optical depths, we naturally find that as $\tau_{\rm IR}$ increases, the peak optical depth for $\gamma-\gamma$ interactions shifts up in value and down in energy, see Fig.~\ref{fig:pairp}.  However, we find that for even very large values for the infrared optical depth, $\tau_{\gamma - \gamma}$ does not rise above unity for energies below 100 GeV.  The $\gamma$-ray energy spectrum measured in \citet{Peng16} extends from $0.1 - 10$ GeV.  Thus, while there is a significant increase in the $\gamma$-ray absorption due to photon-photon interactions above 100 GeV, the spectral region covered by the \textit{Fermi} $\gamma$-ray observations is optically thin.

At present, there are no $\gamma$-ray detections at energies above 10 GeV, only an upper limit at 100 GeV from VERITAS \citep{Fleischhack15}.  However, a future detection by VERITAS or the Cherenkov Telescope Array (CTA) could provide valuable additional constraints on the infrared optical depth and the cosmic ray spectral index.

\begin{figure}
 \subfigure{
  \includegraphics[width=0.90\linewidth]{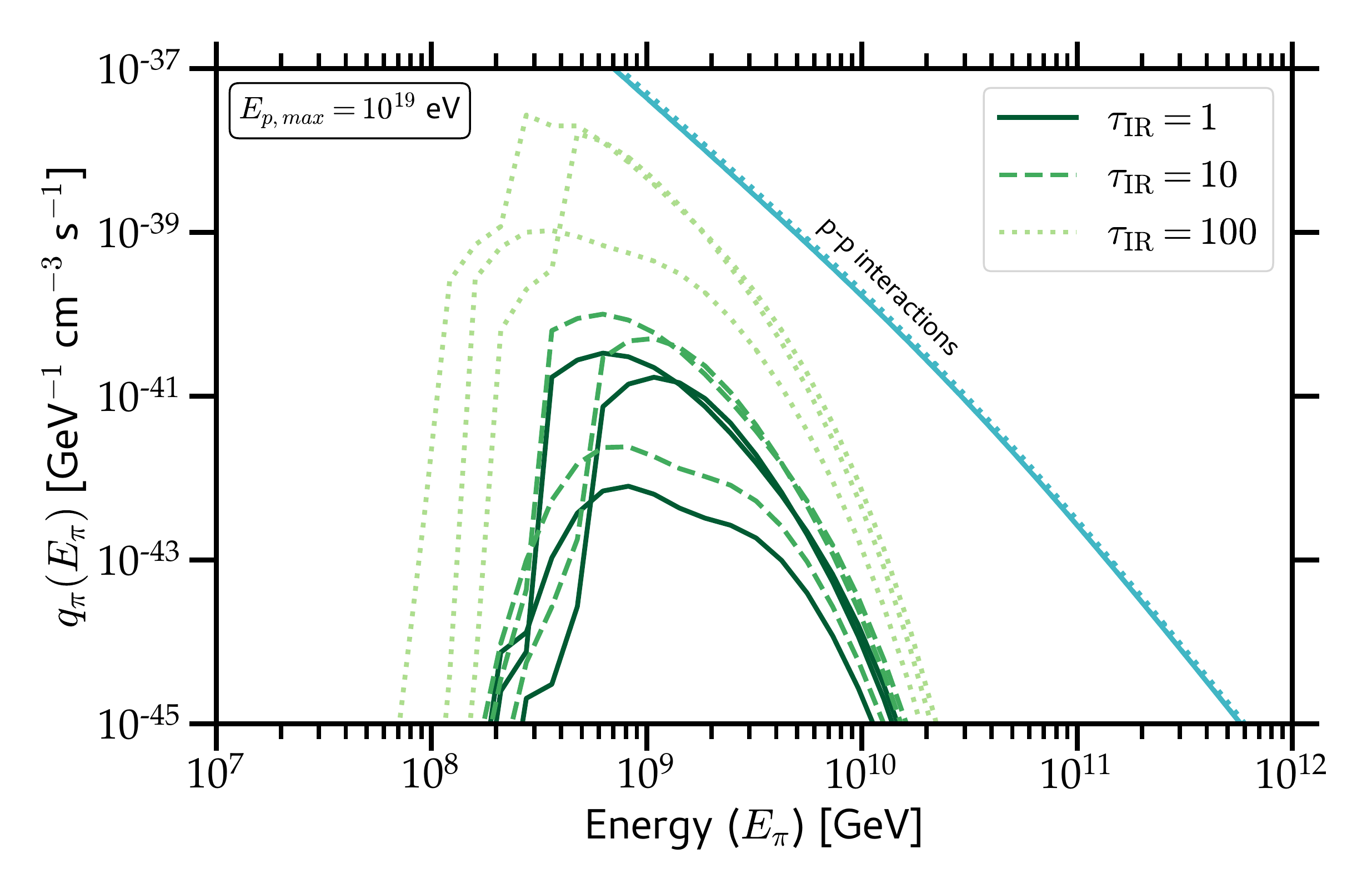}}
 \subfigure{
  \includegraphics[width=0.90\linewidth]{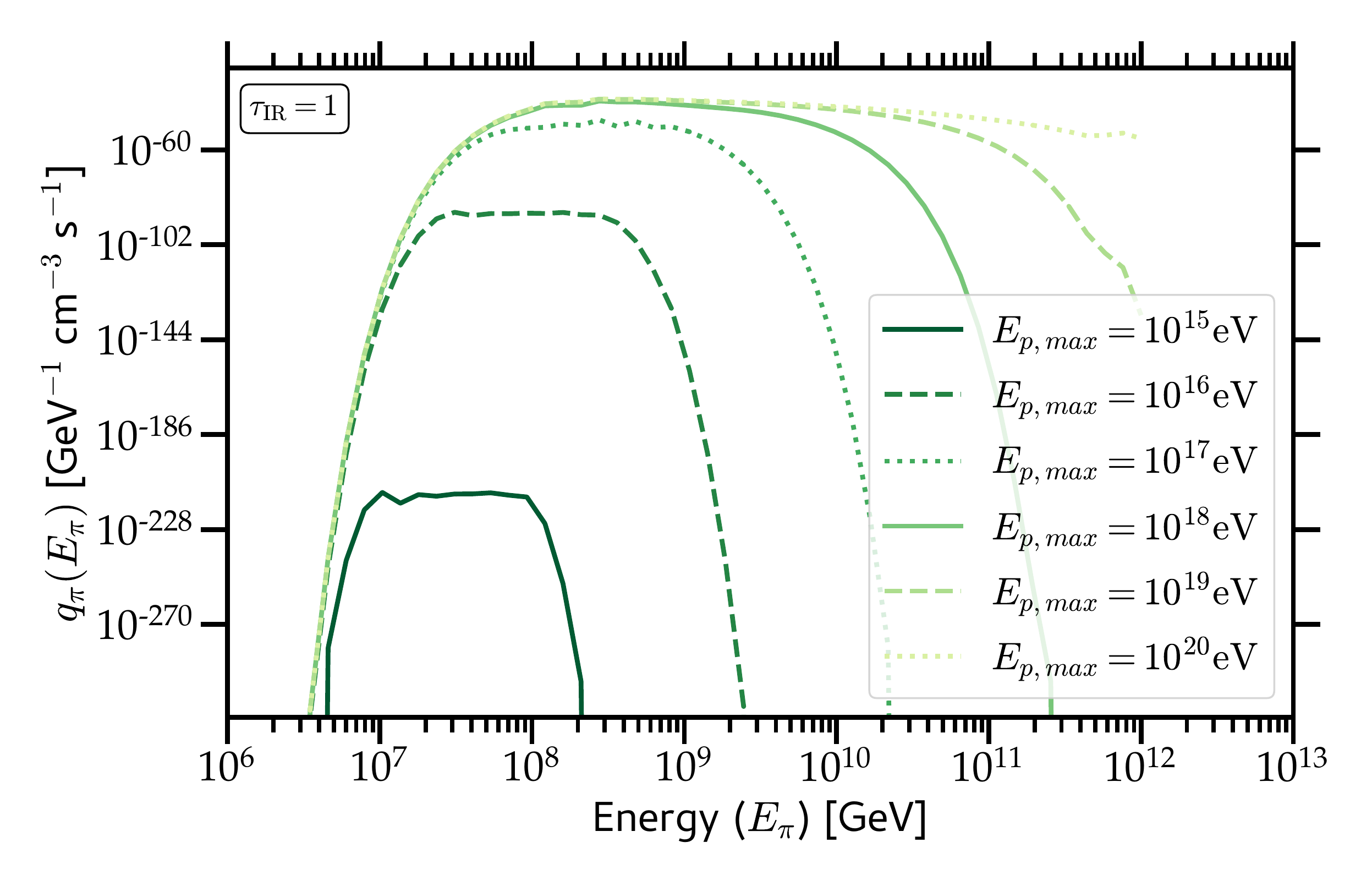}}
\caption{The source function for photopion production is shown for varying infrared optical depths (top) and for varying maximum cosmic ray energies (bottom).  The top figure also includes the source function for pion production from proton-proton interactions (solid top lines).  Maximum cosmic ray energies of greater than $\sim 10^{18}$ are necessary for pion production from photon-proton interactions to even approach the pion production from proton-proton interactions.}
\label{fig:photopion}
\end{figure}
%
%

\subsubsection{Proton-Photon Interactions} \label{sec:pgint}

In an environment with extremely intense radiation fields, pions produced in proton-photon ($p - \gamma$) interactions can be the dominant source of $\gamma$-ray and neutrino production.  Here, we compare the source function for pions from $p - \gamma$ processes against that from $p - p$ processes, see Fig.~\ref{fig:photopion}.  Using the simplified models for photo-hadronic interactions from \citet{Hummer10}, we plot the source function for photopions for a variety of effective infrared optical depths and for varying maximum cosmic ray energies, see Appendix~\ref{sec:appc} for details about the source function.

Examining Fig.~\ref{fig:photopion}, we find that even at large optical depths, the pion flux from $p - p$ interactions is orders of magnitude larger than the pion flux from $p - \gamma$ interactions.  Additionally, the maximum cosmic ray energy must be greater than $10^{18}$ eV for the cosmic ray protons in order for the photopion source function to even approach the source function of pions from proton-proton interactions.

\section{Summary} \label{sec:summary}

As established above and in our previous analysis \citep{YoastHull17a}, there is a significant discrepancy between the cosmic ray energy injection rates derived from the radio spectra versus the $\gamma$-ray spectrum in Arp 220, see Fig.~\ref{fig:spectra}.  Observational constraints on the average ISM density and the magnetic field strength do not allow for adjustment of the radio spectrum through decreases in the synchrotron power spectrum or through increases in the bremsstrahlung energy loss rates, see Fig.~\ref{fig:schematic}.

However, observational diagnostics show that the nuclei of Arp 220 are optically thick over the extended range of the infrared through to millimeter wavelengths. It follows then that the intrinsic infrared energy densities inside the nuclei is significantly larger than the emergent energy densities \citep[e.g.,][]{Wilson14,Scoville17}.  By testing different energy injection scenarios with varying values for an effective infrared optical depth, we have shown above that an optical depth greater than unity can be combined with a large cosmic ray energy injection rate to simultaneously fit the radio and $\gamma$-ray spectra, see Fig.~\ref{fig:gchi} -- \ref{fig:rchi}.

Infrared optical depths calculated from submillimeter wavelength measurements \citep{Wilson14} are within the range of effective optical depths considered acceptable from our $\chi^{2}$ tests.  Further, using these optical depths to calculate the total radiation field energy density, we find equipartition exists in both nuclei between the turbulent energy densities and the radiation field energy densities.

These results are, of course, dependent on several assumptions about the total dense gas mass, dust-to-gas ratio, and additional properties of the dust (e.g., grain size) in Arp 220.  Additionally, as the image of $\gamma$-ray emission from Arp 220 is consistent with that of a point source, we lack any spatial information regarding the $\gamma$-ray flux or the composition of the additional cosmic ray population needed to produce such a $\gamma$-ray flux.  Because of these factors and the complexities related to radiative transfer, we cannot use our results to pinpoint an exact value for the optical depths.

Regardless of the limitations of these models, we have clearly established that the infrared energy density in the nuclei is significantly larger than that in the emergent radiation field.  Thus, models for cosmic ray interactions must properly account for optically thick radiation in calculating radio and $\gamma$-ray emission.  This may have major implications for methods for deriving SFRs from radio emission and calculations of $\gamma$-ray and neutrino emission at high redshift, provided those regions or systems host very compact regions of star formation (i.e., large star-formation surface densities).

\section*{Acknowledgements}

We thank Jay Gallagher, Todd Thompson, and Almog Yalinewich for useful discussions on Arp 220 and on greybodies and radiative transfer.  We also thank Mubdi Rahman for discussions on statistics.
This work was performed in part at the Aspen Center for Physics, which is supported by National Science Foundation grant PHY-1607611. We acknowledge the support of the Natural Sciences and Engineering Research Council of Canada (NSERC). This research was undertaken, in part, thanks to funding from the Canada Research Chairs program.

\bibliography{Bib}

\begin{thebibliography}{}
\makeatletter
\relax
\def\mn@urlcharsother{\let\do\@makeother \do\$\do\&\do\#\do\^\do\_\do\%\do\~}
\def\mn@doi{\begingroup\mn@urlcharsother \@ifnextchar [ {\mn@doi@}
  {\mn@doi@[]}}
\def\mn@doi@[#1]#2{\def\@tempa{#1}\ifx\@tempa\@empty \href
  {http://dx.doi.org/#2} {doi:#2}\else \href {http://dx.doi.org/#2} {#1}\fi
  \endgroup}
\def\mn@eprint#1#2{\mn@eprint@#1:#2::\@nil}
\def\mn@eprint@arXiv#1{\href {http://arxiv.org/abs/#1} {{\tt arXiv:#1}}}
\def\mn@eprint@dblp#1{\href {http://dblp.uni-trier.de/rec/bibtex/#1.xml}
  {dblp:#1}}
\def\mn@eprint@#1:#2:#3:#4\@nil{\def\@tempa {#1}\def\@tempb {#2}\def\@tempc
  {#3}\ifx \@tempc \@empty \let \@tempc \@tempb \let \@tempb \@tempa \fi \ifx
  \@tempb \@empty \def\@tempb {arXiv}\fi \@ifundefined
  {mn@eprint@\@tempb}{\@tempb:\@tempc}{\expandafter \expandafter \csname
  mn@eprint@\@tempb\endcsname \expandafter{\@tempc}}}

\bibitem[\protect\citeauthoryear{{Abdo} et~al.,}{{Abdo} et~al.}{2010a}]{Abdo10}
{Abdo} A.~A.,  et~al., 2010a, \mn@doi [\aap] {10.1051/0004-6361/201015759},
  \href {http://adsabs.harvard.edu/abs/2010A%26A...523L...2A} {523, L2}

\bibitem[\protect\citeauthoryear{{Abdo} et~al.,}{{Abdo}
  et~al.}{2010b}]{Abdo10a}
{Abdo} A.~A.,  et~al., 2010b, \mn@doi [\apjl] {10.1088/2041-8205/709/2/L152},
  \href {http://adsabs.harvard.edu/abs/2010ApJ...709L.152A} {709, L152}

\bibitem[\protect\citeauthoryear{{Acero} et~al.,}{{Acero}
  et~al.}{2015}]{Acero15}
{Acero} F.,  et~al., 2015, \mn@doi [\apjs] {10.1088/0067-0049/218/2/23}, \href
  {http://adsabs.harvard.edu/abs/2015ApJS..218...23A} {218, 23}

\bibitem[\protect\citeauthoryear{{Ackermann} et~al.,}{{Ackermann}
  et~al.}{2012}]{Ackermann12}
{Ackermann} M.,  et~al., 2012, \mn@doi [\apj] {10.1088/0004-637X/755/2/164},
  \href {http://adsabs.harvard.edu/abs/2012ApJ...755..164A} {755, 164}

\bibitem[\protect\citeauthoryear{{Ackermann} et~al.,}{{Ackermann}
  et~al.}{2016}]{Ackermann16}
{Ackermann} M.,  et~al., 2016, \mn@doi [\aap] {10.1051/0004-6361/201526920},
  \href {http://adsabs.harvard.edu/abs/2016A%26A...586A..71A} {586, A71}

\bibitem[\protect\citeauthoryear{{Barcos-Mu{\~n}oz} et~al.,}{{Barcos-Mu{\~n}oz}
  et~al.}{2015}]{Barcos15}
{Barcos-Mu{\~n}oz} L.,  et~al., 2015, \mn@doi [\apj]
  {10.1088/0004-637X/799/1/10}, \href
  {http://adsabs.harvard.edu/abs/2015ApJ...799...10B} {799, 10}

\bibitem[\protect\citeauthoryear{{Blumenthal} \& {Gould}}{{Blumenthal} \&
  {Gould}}{1970}]{Blumenthal70}
{Blumenthal} G.~R.,  {Gould} R.~J.,  1970, \mn@doi [Rev. Mod. Phys.]
  {10.1103/RevModPhys.42.237}, \href
  {http://adsabs.harvard.edu/abs/1970RvMP...42..237B} {42, 237}

\bibitem[\protect\citeauthoryear{{Boettcher}, {Harris}  \&
  {Krawczynski}}{{Boettcher} et~al.}{2012}]{Bottcher12}
{Boettcher} M.,  {Harris} D.~E.,   {Krawczynski} H.,  2012, {Relativistic Jets
  from Active Galactic Nuclei}.
{Wiley}, {Berlin}

\bibitem[\protect\citeauthoryear{{Bolatto}, {Wolfire}  \& {Leroy}}{{Bolatto}
  et~al.}{2013}]{Bolatto13}
{Bolatto} A.~D.,  {Wolfire} M.,   {Leroy} A.~K.,  2013, \mn@doi [\araa]
  {10.1146/annurev-astro-082812-140944}, \href
  {http://adsabs.harvard.edu/abs/2013ARA%26A..51..207B} {51, 207}

\bibitem[\protect\citeauthoryear{{Casey}}{{Casey}}{2012}]{Casey12}
{Casey} C.~M.,  2012, \mn@doi [\mnras] {10.1111/j.1365-2966.2012.21455.x},
  \href {http://adsabs.harvard.edu/abs/2012MNRAS.425.3094C} {425, 3094}

\bibitem[\protect\citeauthoryear{{Condon}, {Cotton}, {Greisen}, {Yin},
  {Perley}, {Taylor}  \& {Broderick}}{{Condon} et~al.}{1998}]{Condon98}
{Condon} J.~J.,  {Cotton} W.~D.,  {Greisen} E.~W.,  {Yin} Q.~F.,  {Perley}
  R.~A.,  {Taylor} G.~B.,   {Broderick} J.~J.,  1998, \mn@doi [\aj]
  {10.1086/300337}, \href {http://adsabs.harvard.edu/abs/1998AJ....115.1693C}
  {115, 1693}

\bibitem[\protect\citeauthoryear{{Condon}, {Cotton}  \& {Broderick}}{{Condon}
  et~al.}{2002}]{Condon02}
{Condon} J.~J.,  {Cotton} W.~D.,   {Broderick} J.~J.,  2002, \mn@doi [\aj]
  {10.1086/341650}, \href {http://adsabs.harvard.edu/abs/2002AJ....124..675C}
  {124, 675}

\bibitem[\protect\citeauthoryear{{Crutcher}}{{Crutcher}}{2012}]{Crutcher12}
{Crutcher} R.~M.,  2012, \mn@doi [\araa] {10.1146/annurev-astro-081811-125514},
  \href {http://adsabs.harvard.edu/abs/2012ARA%26A..50...29C} {50, 29}

\bibitem[\protect\citeauthoryear{{Dennison}, {Balonek}, {Terzian}  \&
  {Balick}}{{Dennison} et~al.}{1975}]{Dennison75}
{Dennison} B.,  {Balonek} T.~J.,  {Terzian} Y.,   {Balick} B.,  1975, \mn@doi
  [\pasp] {10.1086/129725}, \href
  {http://adsabs.harvard.edu/abs/1975PASP...87...83D} {87, 83}

\bibitem[\protect\citeauthoryear{{Dermer} \& {Menon}}{{Dermer} \&
  {Menon}}{2009}]{Dermer09}
{Dermer} C.~D.,  {Menon} G.,  2009, {High Energy Radiation from Black Holes:
  Gamma Rays, Cosmic Rays, and Neutrinos}.
{Princeton Univ. Press}, {Princeton, NJ}

\bibitem[\protect\citeauthoryear{{Downes} \& {Solomon}}{{Downes} \&
  {Solomon}}{1998}]{Downes98}
{Downes} D.,  {Solomon} P.~M.,  1998, \mn@doi [\apj] {10.1086/306339}, \href
  {http://adsabs.harvard.edu/abs/1998ApJ...507..615D} {507, 615}

\bibitem[\protect\citeauthoryear{{Eichmann} \& {Becker Tjus}}{{Eichmann} \&
  {Becker Tjus}}{2016}]{Eichmann16}
{Eichmann} B.,  {Becker Tjus} J.,  2016, \mn@doi [\apj]
  {10.3847/0004-637X/821/2/87}, \href
  {http://adsabs.harvard.edu/abs/2016ApJ...821...87E} {821, 87}

\bibitem[\protect\citeauthoryear{{Elmouttie}, {Haynes}, {Jones}, {Ehle},
  {Beck}, {Harnett}  \& {Wielebinski}}{{Elmouttie} et~al.}{1997}]{Elmouttie97}
{Elmouttie} M.,  {Haynes} R.~F.,  {Jones} K.~L.,  {Ehle} M.,  {Beck} R.,
  {Harnett} J.~I.,   {Wielebinski} R.,  1997, \mn@doi [\mnras]
  {10.1093/mnras/284.4.830}, \href
  {http://adsabs.harvard.edu/abs/1997MNRAS.284..830E} {284, 830}

\bibitem[\protect\citeauthoryear{{Elmouttie}, {Haynes}, {Jones}, {Sadler}  \&
  {Ehle}}{{Elmouttie} et~al.}{1998}]{Elmouttie98}
{Elmouttie} M.,  {Haynes} R.~F.,  {Jones} K.~L.,  {Sadler} E.~M.,   {Ehle} M.,
  1998, \mn@doi [\mnras] {10.1046/j.1365-8711.1998.01592.x}, \href
  {http://adsabs.harvard.edu/abs/1998MNRAS.297.1202E} {297, 1202}

\bibitem[\protect\citeauthoryear{{Fleischhack} \& {VERITAS
  Collaboration}}{{Fleischhack} \& {VERITAS
  Collaboration}}{2015}]{Fleischhack15}
{Fleischhack} H.,  {VERITAS Collaboration} 2015, in 34th International Cosmic
  Ray Conference (ICRC2015). p.~745

\bibitem[\protect\citeauthoryear{{Ghisellini}}{{Ghisellini}}{2013}]{Ghisellini13}
{Ghisellini} G.,  ed. 2013, {Radiative Processes in High Energy Astrophysics}
  Lecture Notes in Physics, Berlin Springer Verlag Vol. 873

\bibitem[\protect\citeauthoryear{{Ginzburg}}{{Ginzburg}}{1969}]{Ginzburg69}
{Ginzburg} V.~L.,  1969, {Elementary Processes for Cosmic Ray Astrophysics}.
{Gordon \& Breach}, {New York}

\bibitem[\protect\citeauthoryear{{Gonz{\'a}lez-Alfonso}, {Smith}, {Fischer}  \&
  {Cernicharo}}{{Gonz{\'a}lez-Alfonso} et~al.}{2004}]{Gonzalez04}
{Gonz{\'a}lez-Alfonso} E.,  {Smith} H.~A.,  {Fischer} J.,   {Cernicharo} J.,
  2004, \mn@doi [\apj] {10.1086/422868}, \href
  {http://adsabs.harvard.edu/abs/2004ApJ...613..247G} {613, 247}

\bibitem[\protect\citeauthoryear{{Griffin}, {Dai}  \& {Thompson}}{{Griffin}
  et~al.}{2016}]{Griffin16}
{Griffin} R.~D.,  {Dai} X.,   {Thompson} T.~A.,  2016, \mn@doi [\apjl]
  {10.3847/2041-8205/823/1/L17}, \href
  {http://adsabs.harvard.edu/abs/2016ApJ...823L..17G} {823, L17}

\bibitem[\protect\citeauthoryear{{Hayashida} et~al.,}{{Hayashida}
  et~al.}{2013}]{Hayashida13}
{Hayashida} M.,  et~al., 2013, \mn@doi [\apj] {10.1088/0004-637X/779/2/131},
  \href {http://adsabs.harvard.edu/abs/2013ApJ...779..131H} {779, 131}

\bibitem[\protect\citeauthoryear{{Helou} \& {Walker}}{{Helou} \&
  {Walker}}{1988}]{Helou88}
{Helou} G.,  {Walker} D.~W.,  eds, 1988, {Infrared astronomical satellite
  (IRAS) catalogs and atlases. Volume 7: The small scale structure catalog}
  Vol. 7

\bibitem[\protect\citeauthoryear{{Helou}, {Soifer}  \&
  {Rowan-Robinson}}{{Helou} et~al.}{1985}]{Helou85}
{Helou} G.,  {Soifer} B.~T.,   {Rowan-Robinson} M.,  1985, \mn@doi [\apjl]
  {10.1086/184556}, \href {http://adsabs.harvard.edu/abs/1985ApJ...298L...7H}
  {298, L7}

\bibitem[\protect\citeauthoryear{{H{\"u}mmer}, {R{\"u}ger}, {Spanier}  \&
  {Winter}}{{H{\"u}mmer} et~al.}{2010}]{Hummer10}
{H{\"u}mmer} S.,  {R{\"u}ger} M.,  {Spanier} F.,   {Winter} W.,  2010, \mn@doi
  [\apj] {10.1088/0004-637X/721/1/630}, \href
  {http://adsabs.harvard.edu/abs/2010ApJ...721..630H} {721, 630}

\bibitem[\protect\citeauthoryear{{Kelner}, {Aharonian}  \& {Bugayov}}{{Kelner}
  et~al.}{2006}]{Kelner06}
{Kelner} S.~R.,  {Aharonian} F.~A.,   {Bugayov} V.~V.,  2006, \mn@doi [\prd]
  {10.1103/PhysRevD.74.034018}, \href
  {http://adsabs.harvard.edu/abs/2006PhRvD..74c4018K} {74, 034018}

\bibitem[\protect\citeauthoryear{{Lacki} \& {Thompson}}{{Lacki} \&
  {Thompson}}{2013}]{Lacki13b}
{Lacki} B.~C.,  {Thompson} T.~A.,  2013, \mn@doi [\apj]
  {10.1088/0004-637X/762/1/29}, \href
  {http://adsabs.harvard.edu/abs/2013ApJ...762...29L} {762, 29}

\bibitem[\protect\citeauthoryear{{Lacki}, {Thompson}  \& {Quataert}}{{Lacki}
  et~al.}{2010}]{Lacki10}
{Lacki} B.~C.,  {Thompson} T.~A.,   {Quataert} E.,  2010, \mn@doi [\apj]
  {10.1088/0004-637X/717/1/1}, \href
  {http://adsabs.harvard.edu/abs/2010ApJ...717....1L} {717, 1}

\bibitem[\protect\citeauthoryear{{Lacki}, {Thompson}, {Quataert}, {Loeb}  \&
  {Waxman}}{{Lacki} et~al.}{2011}]{Lacki11}
{Lacki} B.~C.,  {Thompson} T.~A.,  {Quataert} E.,  {Loeb} A.,   {Waxman} E.,
  2011, \mn@doi [\apj] {10.1088/0004-637X/734/2/107}, \href
  {http://adsabs.harvard.edu/abs/2011ApJ...734..107L} {734, 107}

\bibitem[\protect\citeauthoryear{{Lenain}, {Ricci}, {T{\"u}rler}, {Dorner}  \&
  {Walter}}{{Lenain} et~al.}{2010}]{Lenain10}
{Lenain} J.-P.,  {Ricci} C.,  {T{\"u}rler} M.,  {Dorner} D.,   {Walter} R.,
  2010, \mn@doi [\aap] {10.1051/0004-6361/201015644}, \href
  {http://adsabs.harvard.edu/abs/2010A%26A...524A..72L} {524, A72}

\bibitem[\protect\citeauthoryear{{Lisenfeld}, {Voelk}  \& {Xu}}{{Lisenfeld}
  et~al.}{1996}]{Lisenfeld96a}
{Lisenfeld} U.,  {Voelk} H.~J.,   {Xu} C.,  1996, \aap, \href
  {http://adsabs.harvard.edu/abs/1996A%26A...306..677L} {306, 677}

\bibitem[\protect\citeauthoryear{{Lopez}, {Auchettl}, {Linden}, {Bolatto},
  {Thompson}  \& {Ramirez-Ruiz}}{{Lopez} et~al.}{2018}]{Lopez18}
{Lopez} L.~A.,  {Auchettl} K.,  {Linden} T.,  {Bolatto} A.~D.,  {Thompson}
  T.~A.,   {Ramirez-Ruiz} E.,  2018, \apj, \href
  {http://adsabs.harvard.edu/abs/2018ApJ...867...44L} {867}

\bibitem[\protect\citeauthoryear{{Martin}}{{Martin}}{2014}]{Pohl14}
{Martin} P.,  2014, \mn@doi [\aap] {10.1051/0004-6361/201323329}, \href
  {http://adsabs.harvard.edu/abs/2014A%26A...564A..61M} {564, A61}

\bibitem[\protect\citeauthoryear{{McBride}, {Robishaw}, {Heiles}, {Bower}  \&
  {Sarma}}{{McBride} et~al.}{2015}]{McBride15}
{McBride} J.,  {Robishaw} T.,  {Heiles} C.,  {Bower} G.~C.,   {Sarma} A.~P.,
  2015, \mn@doi [\mnras] {10.1093/mnras/stu2489}, \href
  {http://adsabs.harvard.edu/abs/2015MNRAS.447.1103M} {447, 1103}

\bibitem[\protect\citeauthoryear{{M{\"u}cke}, {Engel}, {Rachen}, {Protheroe}
  \& {Stanev}}{{M{\"u}cke} et~al.}{2000}]{Mucke00}
{M{\"u}cke} A.,  {Engel} R.,  {Rachen} J.~P.,  {Protheroe} R.~J.,   {Stanev}
  T.,  2000, \mn@doi [Computer Physics Communications]
  {10.1016/S0010-4655(99)00446-4}, \href
  {http://adsabs.harvard.edu/abs/2000CoPhC.124..290M} {124, 290}

\bibitem[\protect\citeauthoryear{{Peng}, {Wang}, {Liu}, {Tang}  \&
  {Wang}}{{Peng} et~al.}{2016}]{Peng16}
{Peng} F.-K.,  {Wang} X.-Y.,  {Liu} R.-Y.,  {Tang} Q.-W.,   {Wang} J.-F.,
  2016, \mn@doi [\apjl] {10.3847/2041-8205/821/2/L20}, \href
  {http://adsabs.harvard.edu/abs/2016ApJ...821L..20P} {821, L20}

\bibitem[\protect\citeauthoryear{{Peretti}, {Blasi}, {Aharonian}  \&
  {Morlino}}{{Peretti} et~al.}{2018}]{Peretti18}
{Peretti} E.,  {Blasi} P.,  {Aharonian} F.,   {Morlino} G.,  2018, arXiv
  e-prints, \href {http://adsabs.harvard.edu/abs/2018arXiv181201996P} {}

\bibitem[\protect\citeauthoryear{{Pfrommer}, {Pakmor}, {Simpson}  \&
  {Springel}}{{Pfrommer} et~al.}{2017}]{Pfrommer17}
{Pfrommer} C.,  {Pakmor} R.,  {Simpson} C.~M.,   {Springel} V.,  2017, \mn@doi
  [\apjl] {10.3847/2041-8213/aa8bb1}, \href
  {http://adsabs.harvard.edu/abs/2017ApJ...847L..13P} {847, L13}

\bibitem[\protect\citeauthoryear{{Rangwala} et~al.,}{{Rangwala}
  et~al.}{2011}]{Rangwala11}
{Rangwala} N.,  et~al., 2011, \mn@doi [\apj] {10.1088/0004-637X/743/1/94},
  \href {http://adsabs.harvard.edu/abs/2011ApJ...743...94R} {743, 94}

\bibitem[\protect\citeauthoryear{{Rodr{\'{\i}}guez-Rico}, {Goss}, {Viallefond},
  {Zhao}, {G{\'o}mez}  \& {Anantharamaiah}}{{Rodr{\'{\i}}guez-Rico}
  et~al.}{2005}]{Rodriguez05}
{Rodr{\'{\i}}guez-Rico} C.~A.,  {Goss} W.~M.,  {Viallefond} F.,  {Zhao} J.-H.,
  {G{\'o}mez} Y.,   {Anantharamaiah} K.~R.,  2005, \mn@doi [\apj]
  {10.1086/444491}, \href {http://adsabs.harvard.edu/abs/2005ApJ...633..198R}
  {633, 198}

\bibitem[\protect\citeauthoryear{{Rojas-Bravo} \& {Araya}}{{Rojas-Bravo} \&
  {Araya}}{2016}]{Rojas16}
{Rojas-Bravo} C.,  {Araya} M.,  2016, \mn@doi [\mnras] {10.1093/mnras/stw2059},
  \href {http://adsabs.harvard.edu/abs/2016MNRAS.463.1068R} {463, 1068}

\bibitem[\protect\citeauthoryear{{Rybicki} \& {Lightman}}{{Rybicki} \&
  {Lightman}}{1979}]{RL79}
{Rybicki} G.~B.,  {Lightman} A.~P.,  1979, {Radiative Processes in
  Astrophysics}.
{Wiley-Interscience}, {New York}

\bibitem[\protect\citeauthoryear{{Sakamoto} et~al.,}{{Sakamoto}
  et~al.}{2017}]{Sakamoto17}
{Sakamoto} K.,  et~al., 2017, \mn@doi [\apj] {10.3847/1538-4357/aa8f4b}, \href
  {http://adsabs.harvard.edu/abs/2017ApJ...849...14S} {849, 14}

\bibitem[\protect\citeauthoryear{{Sanders} \& {Mirabel}}{{Sanders} \&
  {Mirabel}}{1996}]{Sanders96}
{Sanders} D.~B.,  {Mirabel} I.~F.,  1996, \mn@doi [\araa]
  {10.1146/annurev.astro.34.1.749}, \href
  {http://adsabs.harvard.edu/abs/1996ARA%26A..34..749S} {34, 749}

\bibitem[\protect\citeauthoryear{{Sanders}, {Mazzarella}, {Kim}, {Surace}  \&
  {Soifer}}{{Sanders} et~al.}{2003}]{Sanders03}
{Sanders} D.~B.,  {Mazzarella} J.~M.,  {Kim} D.-C.,  {Surace} J.~A.,   {Soifer}
  B.~T.,  2003, \mn@doi [\aj] {10.1086/376841}, \href
  {http://adsabs.harvard.edu/abs/2003AJ....126.1607S} {126, 1607}

\bibitem[\protect\citeauthoryear{{Schlickeiser}}{{Schlickeiser}}{2002}]{Schlick02}
{Schlickeiser} R.,  2002, {Cosmic Ray Astrophysics}.
{Springer}, {Berlin}

\bibitem[\protect\citeauthoryear{{Schlickeiser} \& {Ruppel}}{{Schlickeiser} \&
  {Ruppel}}{2010}]{Schlick10}
{Schlickeiser} R.,  {Ruppel} J.,  2010, \mn@doi [New J. of Phys.]
  {10.1088/1367-2630/12/3/033044}, \href
  {http://adsabs.harvard.edu/abs/2010NJPh...12c3044S} {12, 033044}

\bibitem[\protect\citeauthoryear{{Sch{\"o}neberg}, {Becker Tjus}  \&
  {Schuppan}}{{Sch{\"o}neberg} et~al.}{2013}]{Schoneberg13}
{Sch{\"o}neberg} S.,  {Becker Tjus} J.,   {Schuppan} F.,  2013, in {Torres}
  D.~F.,  {Reimer} O.,  eds,  Astrophysics and Space Science Proceedings Vol.
  34, Cosmic Rays in Star-Forming Environments. p.~299 (\mn@eprint {arXiv}
  {1209.1196}), \mn@doi{10.1007/978-3-642-35410-6_21}

\bibitem[\protect\citeauthoryear{{Scoville} et~al.,}{{Scoville}
  et~al.}{2017}]{Scoville17}
{Scoville} N.,  et~al., 2017, \mn@doi [\apj] {10.3847/1538-4357/836/1/66},
  \href {http://adsabs.harvard.edu/abs/2017ApJ...836...66S} {836, 66}

\bibitem[\protect\citeauthoryear{{Soifer}, {Boehmer}, {Neugebauer}  \&
  {Sanders}}{{Soifer} et~al.}{1989}]{Soifer89}
{Soifer} B.~T.,  {Boehmer} L.,  {Neugebauer} G.,   {Sanders} D.~B.,  1989,
  \mn@doi [\aj] {10.1086/115178}, \href
  {http://adsabs.harvard.edu/abs/1989AJ.....98..766S} {98, 766}

\bibitem[\protect\citeauthoryear{{Stecker}}{{Stecker}}{1971}]{Stecker71}
{Stecker} F.~W.,  1971, NASA Special Publication, \href
  {http://adsabs.harvard.edu/abs/1971NASSP.249.....S} {249}

\bibitem[\protect\citeauthoryear{{Tang}, {Wang}  \& {Tam}}{{Tang}
  et~al.}{2014}]{Tang14}
{Tang} Q.-W.,  {Wang} X.-Y.,   {Tam} P.-H.~T.,  2014, \mn@doi [\apj]
  {10.1088/0004-637X/794/1/26}, \href
  {http://adsabs.harvard.edu/abs/2014ApJ...794...26T} {794, 26}

\bibitem[\protect\citeauthoryear{{Thompson}, {Quataert}, {Waxman}, {Murray}  \&
  {Martin}}{{Thompson} et~al.}{2006}]{Thompson06}
{Thompson} T.~A.,  {Quataert} E.,  {Waxman} E.,  {Murray} N.,   {Martin} C.~L.,
   2006, \mn@doi [\apj] {10.1086/504035}, \href
  {http://adsabs.harvard.edu/abs/2006ApJ...645..186T} {645, 186}

\bibitem[\protect\citeauthoryear{{Thompson}, {Quataert}  \&
  {Waxman}}{{Thompson} et~al.}{2007}]{Thompson07}
{Thompson} T.~A.,  {Quataert} E.,   {Waxman} E.,  2007, \mn@doi [\apj]
  {10.1086/509068}, \href {http://adsabs.harvard.edu/abs/2007ApJ...654..219T}
  {654, 219}

\bibitem[\protect\citeauthoryear{{Torres}}{{Torres}}{2004}]{Torres04}
{Torres} D.~F.,  2004, \mn@doi [\apj] {10.1086/425415}, \href
  {http://adsabs.harvard.edu/abs/2004ApJ...617..966T} {617, 966}

\bibitem[\protect\citeauthoryear{{Varenius} et~al.,}{{Varenius}
  et~al.}{2016}]{Varenius16}
{Varenius} E.,  et~al., 2016, \mn@doi [\aap] {10.1051/0004-6361/201628702},
  \href {http://adsabs.harvard.edu/abs/2016A%26A...593A..86V} {593, A86}

\bibitem[\protect\citeauthoryear{{Voelk}}{{Voelk}}{1989}]{Volk89}
{Voelk} H.~J.,  1989, \aap, \href
  {http://adsabs.harvard.edu/abs/1989A%26A...218...67V} {218, 67}

\bibitem[\protect\citeauthoryear{{Williams} \& {Bower}}{{Williams} \&
  {Bower}}{2010}]{Williams10}
{Williams} P.~K.~G.,  {Bower} G.~C.,  2010, \mn@doi [\apj]
  {10.1088/0004-637X/710/2/1462}, \href
  {http://adsabs.harvard.edu/abs/2010ApJ...710.1462W} {710, 1462}

\bibitem[\protect\citeauthoryear{{Wilson}, {Rangwala}, {Glenn}, {Maloney},
  {Spinoglio}  \& {Pereira-Santaella}}{{Wilson} et~al.}{2014}]{Wilson14}
{Wilson} C.~D.,  {Rangwala} N.,  {Glenn} J.,  {Maloney} P.~R.,  {Spinoglio} L.,
    {Pereira-Santaella} M.,  2014, \mn@doi [\apjl]
  {10.1088/2041-8205/789/2/L36}, \href
  {http://adsabs.harvard.edu/abs/2014ApJ...789L..36W} {789, L36}

\bibitem[\protect\citeauthoryear{{Wojaczy{\'n}ski}, {Nied{\'z}wiecki}, {Xie}
  \& {Szanecki}}{{Wojaczy{\'n}ski} et~al.}{2015}]{Wojaczynski15}
{Wojaczy{\'n}ski} R.,  {Nied{\'z}wiecki} A.,  {Xie} F.-G.,   {Szanecki} M.,
  2015, \mn@doi [\aap] {10.1051/0004-6361/201526621}, \href
  {http://adsabs.harvard.edu/abs/2015A%26A...584A..20W} {584, A20}

\bibitem[\protect\citeauthoryear{{Xu} \& {Helou}}{{Xu} \& {Helou}}{1996}]{Xu96}
{Xu} C.,  {Helou} G.,  1996, \mn@doi [\apj] {10.1086/176636}, \href
  {http://adsabs.harvard.edu/abs/1996ApJ...456..152X} {456, 152}

\bibitem[\protect\citeauthoryear{{Yoast-Hull}, {Everett}, {Gallagher}  \&
  {Zweibel}}{{Yoast-Hull} et~al.}{2013}]{YoastHull13}
{Yoast-Hull} T.~M.,  {Everett} J.~E.,  {Gallagher} III J.~S.,   {Zweibel}
  E.~G.,  2013, \mn@doi [\apj] {10.1088/0004-637X/768/1/53}, \href
  {http://adsabs.harvard.edu/abs/2013ApJ...768...53Y} {768, 53}

\bibitem[\protect\citeauthoryear{{Yoast-Hull}, {Gallagher}  \&
  {Zweibel}}{{Yoast-Hull} et~al.}{2014}]{YoastHull14a}
{Yoast-Hull} T.~M.,  {Gallagher} III J.~S.,   {Zweibel} E.~G.,  2014, \mn@doi
  [\apj] {10.1088/0004-637X/790/2/86}, \href
  {http://adsabs.harvard.edu/abs/2014ApJ...790...86Y} {790, 86}

\bibitem[\protect\citeauthoryear{{Yoast-Hull}, {Gallagher}  \&
  {Zweibel}}{{Yoast-Hull} et~al.}{2015}]{YoastHull15}
{Yoast-Hull} T.~M.,  {Gallagher} J.~S.,   {Zweibel} E.~G.,  2015, \mn@doi
  [\mnras] {10.1093/mnras/stv1525}, \href
  {http://adsabs.harvard.edu/abs/2015MNRAS.453..222Y} {453, 222}

\bibitem[\protect\citeauthoryear{{Yoast-Hull}, {Gallagher}  \&
  {Zweibel}}{{Yoast-Hull} et~al.}{2016}]{YoastHull16}
{Yoast-Hull} T.~M.,  {Gallagher} J.~S.,   {Zweibel} E.~G.,  2016, \mn@doi
  [\mnras] {10.1093/mnrasl/slv195}, \href
  {http://adsabs.harvard.edu/abs/2016MNRAS.457L..29Y} {457, L29}

\bibitem[\protect\citeauthoryear{{Yoast-Hull}, {Gallagher}, {Aalto}  \&
  {Varenius}}{{Yoast-Hull} et~al.}{2017}]{YoastHull17a}
{Yoast-Hull} T.~M.,  {Gallagher} III J.~S.,  {Aalto} S.,   {Varenius} E.,
  2017, \mn@doi [\mnras] {10.1093/mnrasl/slx054}, \href
  {http://adsabs.harvard.edu/abs/2017MNRAS.469L..89Y} {469, L89}

\bibitem[\protect\citeauthoryear{{Yun}, {Reddy}  \& {Condon}}{{Yun}
  et~al.}{2001}]{Yun01}
{Yun} M.~S.,  {Reddy} N.~A.,   {Condon} J.~J.,  2001, \mn@doi [\apj]
  {10.1086/323145}, \href {http://adsabs.harvard.edu/abs/2001ApJ...554..803Y}
  {554, 803}

\bibitem[\protect\citeauthoryear{{de Jong}, {Klein}, {Wielebinski}  \&
  {Wunderlich}}{{de Jong} et~al.}{1985}]{deJong85}
{de Jong} T.,  {Klein} U.,  {Wielebinski} R.,   {Wunderlich} E.,  1985, \aap,
  \href {http://adsabs.harvard.edu/abs/1985A%26A...147L...6D} {147, L6}

\makeatother
\end{thebibliography}
\bibliographystyle{mnras}

\appendix

\section{Nearby Galaxies} \label{sec:appa}

\subsection{Regression Analysis}

\begin{figure}
 \subfigure{
  \includegraphics[width=0.84\linewidth]{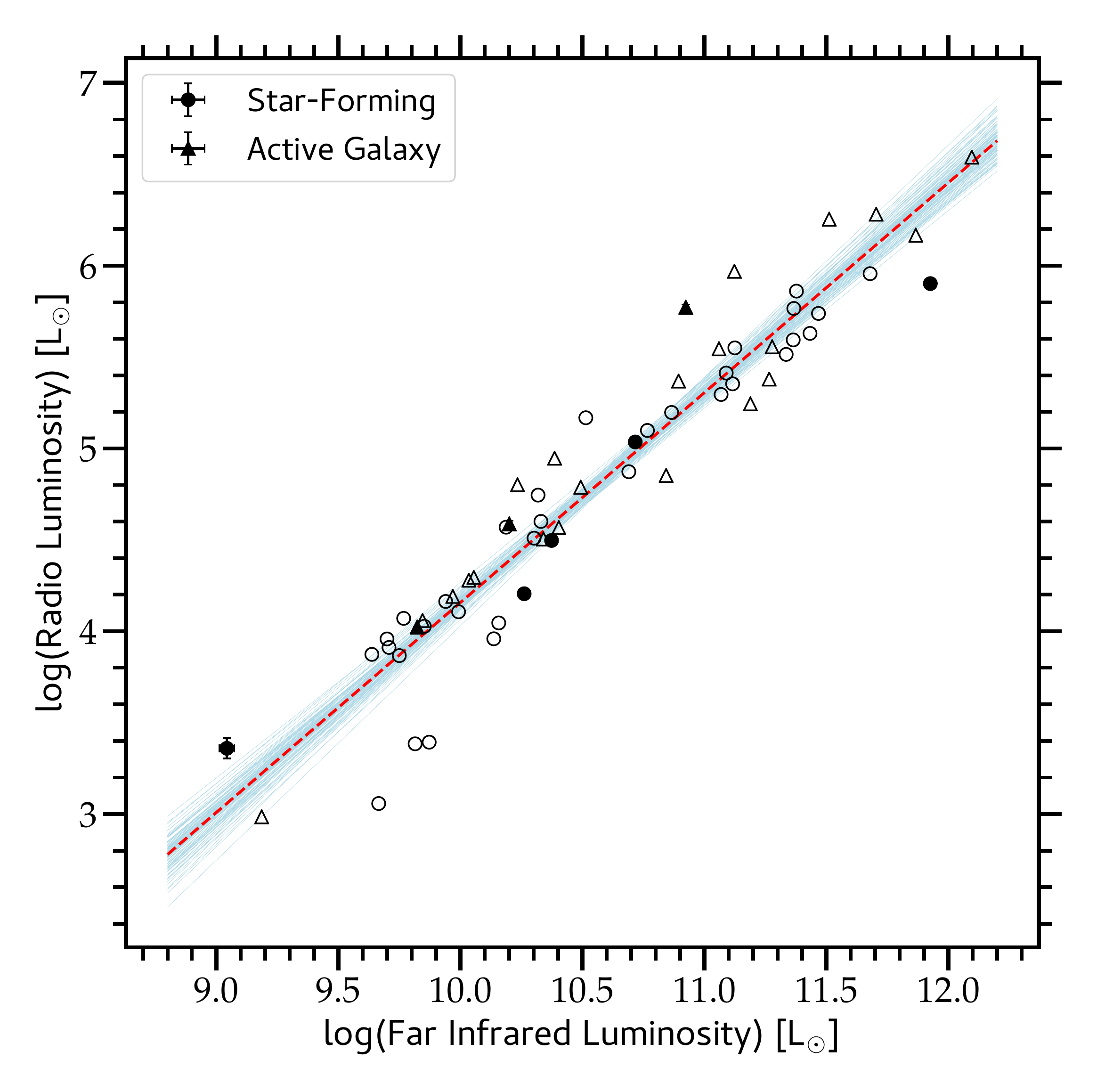}}
 \subfigure{
  \includegraphics[width=0.84\linewidth]{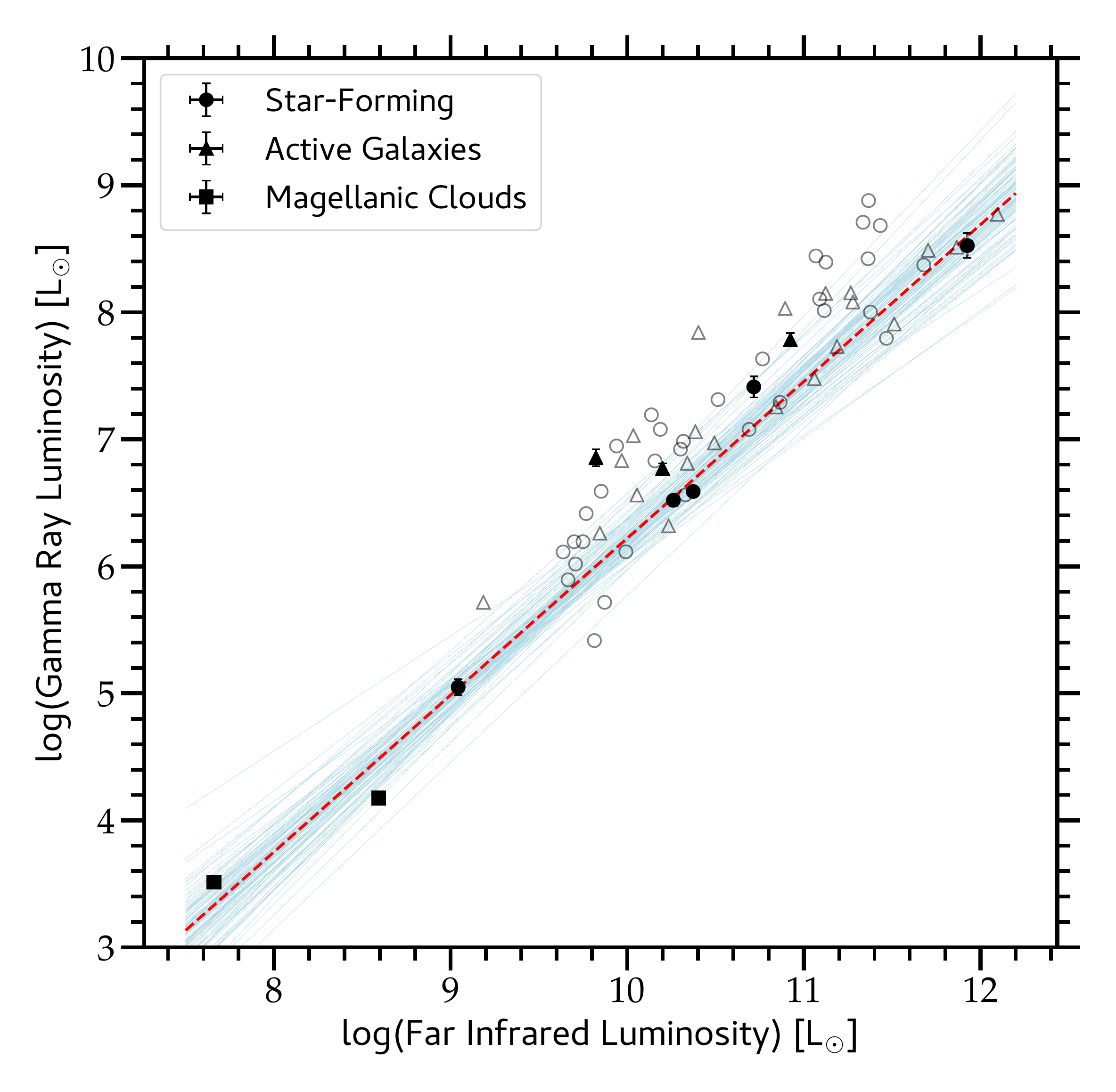}}
 \subfigure{
  \includegraphics[width=0.84\linewidth]{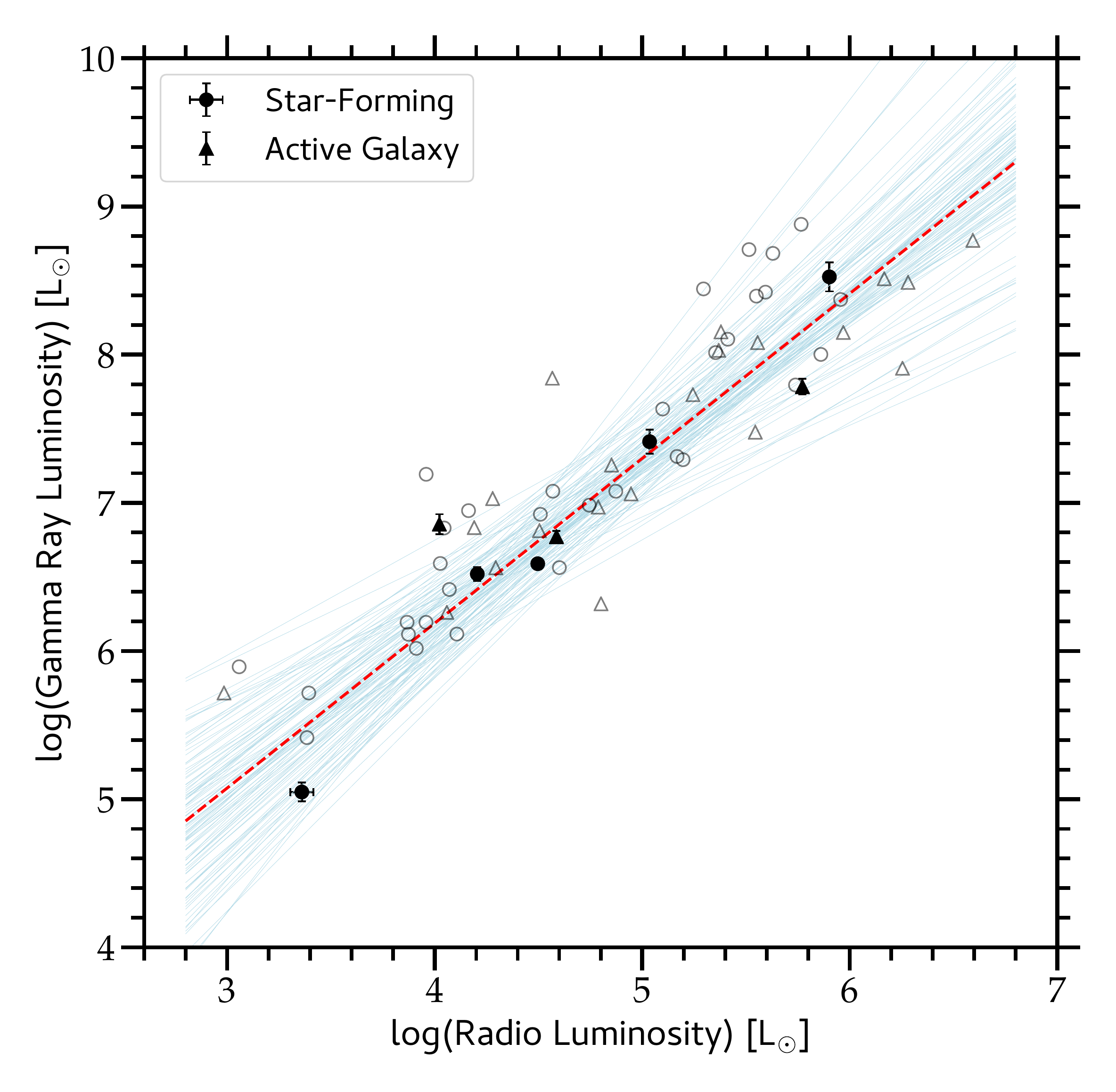}}
\caption{For nearby star-forming galaxies, we show radio versus infrared luminosity (top), gamma-ray versus infrared luminosity (middle), and gamma-ray versus radio luminosity (bottom).  Galaxies with $\gamma$-ray detections by \textit{Fermi} are shown with filled symbols and open gray symbols indicate galaxies with $\gamma$-ray flux upper limits.  Posterior predictive regression lines (100 of 10,000) are shown in solid light blue, with the mean trend line shown in dashed red.}
\label{fig:fluxflux}
\end{figure}

The correlation between the diffuse FIR and radio emission in star-forming galaxies is well-established.  Recent detections of star-forming galaxies in $\gamma$-rays have shown a second correlation between the diffuse FIR and $\gamma$-ray emission, see Fig.~\ref{fig:fluxflux}.  Additionally, we show here that there is a third correlation between the diffuse $\gamma$-ray and radio emission in star-forming galaxies.  

For the galaxies listed in Tables~\ref{tab:data1} \&~\ref{tab:data2}, we perform a Bayesian linear regression analysis on the radio and FIR luminosities.  Assuming normal priors, the mean slope and intercept are $m_{A} = 1.148 \pm 0.047$ and $b_{A} = -7.324 \pm 0.501$, respectively, with a standard deviation of $\sigma_{A} = 0.258$.  The slope derived here is slightly different from the best-fitting slope in \citet{Yun01}; this is expected due to the difference in samples.  Here, we are concentrating on a small sample ($N \sim 60$) of nearby IR-bright galaxies, while general studies of the FIR--radio correlation have sample sizes of $\gtrsim 1800$.

For the galaxies listed in only Table~\ref{tab:data1}, we also perform a Bayesian linear regression analysis on the FIR and $\gamma$-ray luminosities and the radio and $\gamma$-ray luminosities.  Again assuming normal priors, we obtain mean slopes and intercepts of $m_{B} = 1.235 \pm 0.093$ and $b_{B} = -6.125 \pm 0.920$ with a standard deviation of $\sigma_{B} = 0.297$ for the FIR and $\gamma$-ray luminosities and $m_{C} = 1.112 \pm 0.223$ and $b_{C} = 1.739 \pm 1.061$ with a standard deviation of $\sigma_{C} = 0.481$ for the radio and $\gamma$-ray luminosities.

In Fig.~\ref{fig:fluxflux}, we show FIR, $\gamma$-ray, and radio luminosities for 60 nearby galaxies.  Galaxies with $\gamma$-ray detections are shown in filled symbols and those with only $\gamma$-ray upper limits are shown in open symbols.  Pure star-forming galaxies are denoted with circular symbols and active galaxies are denoted with triangular symbols.

For each analysis, we ran 10,000 samples with a burn sample of 1000.  Modeling for the radio and FIR luminosities converged.  However, there were $\sim 2000$ and $\sim 200$ divergences for the FIR and $\gamma$-ray luminosities and the radio and $\gamma$-ray luminosities, respectively.  With an extremely small sample size of 8 galaxies and probable outliers (e.g., Circinus), there is little to be done to create convergence.  Thus, we see a significantly larger spread in the posterior models for the bottom two figures in Fig.~\ref{fig:fluxflux}.

\subsection{Multi-Wavelength Observations}

The total far-infrared fluxes for each galaxy are calculated such that
\begin{equation}
F_{\rm FIR} = 1.26 \times 10^{-11} (2.58 S_{60} + S_{100}),
\end{equation}
where $F_{\rm FIR}$ is given in units of erg cm$^{-2}$ s$^{-1}$ \citep[][and references therein]{Sanders96}.  Infrared data are taken from the \textit{IRAS} Revised Bright Galaxy Sample \citep{Sanders03} and radio data are taken from the NRAO VLA Sky Survey \citep[NVSS;][]{Condon98}. $\gamma$-ray data in Table~\ref{tab:data1} are taken from the Third Fermi Source Catalog \citep{Acero15} and from \citet{Lopez18} and \citet{Ackermann16} for the SMC and the LMC, respectively.  Upper-limits for $\gamma$-ray fluxes in Table~\ref{tab:data2} are taken from \citet{Rojas16}. Exceptions:

\textit{M31 (NGC 0224)}: The infrared fluxes are taken from \citet{Xu96}, and the radio flux is taken from \citet{Dennison75}.

\textit{NGC 1144}: The infrared fluxes are taken from the \textit{IRAS} Bright Galaxy Sample \citep{Soifer89}.

\textit{NGC 3627}: The radio flux used is the sum of both NVSS components associated with the galaxy, following \citet{Condon02}.

\textit{NGC 4945}: The radio flux is taken from \citet{Elmouttie97}.

\textit{NGC 5055}: The radio flux is taken from \citet{Condon02} and is inconsistent with the flux listed in the NVSS catalog.

\textit{M51 (NGC 5194)}: The radio flux is taken from \citet{Condon02} and references therein.

\textit{Circinus}: The infrared fluxes are taken from the \textit{IRAS} Point Source Catalog \citep{Helou88}, and the radio flux is taken from \citet{Elmouttie98}.

\textit{NGC 7331}: The radio flux used is the sum of both NVSS components associated with the galaxy, following \citet{Condon02}.

\begin{table*} 
\begin{minipage}{170mm}
\begin{center}
\caption{Data for Star-Forming Galaxies with Gamma-Ray Detections}
\begin{tabular}{ccccccccc}
\hline
Galaxy & $S_{60~\micron}$ & $\sigma_{60~\micron}$ & $S_{100~\micron}$ & $\sigma_{100~\micron}$ & $S_{1.4 ~ \rm GHz}$ & $\sigma_{1.4 ~ \rm GHz}$ & $F_{0.1-100 ~ \rm GeV}$ & $\sigma_{0.1-100 ~ \rm GeV}$\\
 & (Jy) & (mJy) & (Jy) & (mJy) & (mJy) & (mJy) & (erg cm$^{-2}$ s$^{-1}$) & (erg cm$^{-2}$ s$^{-1}$) \\
\hline
NGC 0224 & 595 & $2.8 \times 10^{4}$ & 3073 & $3.16 \times 10^{5}$ & 8600 & 1100 & $5.88 \times 10^{-12}$ & $0.863 \times 10^{-12}$ \\
NGC 0253 & 968 & 65 & 1290 & 644 & 2990 & 114 & $8.65 \times 10^{-12}$ & $9.22 \times 10^{-13}$ \\
NGC 1068 & 196 & 108 & 257 & 228 & 4850 & 170.4 & $6.97 \times 10^{-12}$ & $8.55 \times 10^{-13}$ \\
NGC 2146 & 147 & 63 & 194 & 240 & 1070 & 39.9 & $3.58 \times 10^{-12}$ & $6.75 \times 10^{-13}$ \\
NGC 3034 & 1480 & 74 & 1370 & 694 & 6200 & 209 & $1.07 \times 10^{-11}$ & $8.37 \times 10^{-13}$ \\
NGC 4945 & 625 & 133 & 1330 & 797 & 6450 & 250 & $1.38 \times 10^{-11}$ & $1.24 \times 10^{-12}$ \\
CIRCINUS & 249 & 24.9 & 316 & 31.6 & 1370 & -- & $1.30 \times 10^{-11}$ & $2.01 \times 10^{-12}$ \\
ARP 220 & 104 & 112 & 115 & 138 & 327 & 9.8 & $1.92 \times 10^{-12}$ & $4.30 \times 10^{-13}$ \\
SMC &  $6.69 \times 10^{3}$ & -- & $1.50 \times 10^{4}$ & -- & -- & -- & $2.90 \times 10^{-11}$ & $1.0 \times 10^{-12}$ \\
LMC & $8.29 \times 10^{4}$ & -- & $1.85 \times 10^{5}$ & -- & -- & -- & $2.24 \times 10^{-10}$ & -- \\
\hline
\multicolumn{9}{l}{\textit{Notes:} NGC 0224 = M31, NGC 3034 = M82.}\\
\label{tab:data1}
\end{tabular}
\end{center}
\end{minipage}
\end{table*}
\begin{table*} 
\begin{minipage}{170mm}
\begin{center}
\caption{Data for Star-Forming Galaxies without Gamma-Ray Detections}
\begin{tabular}{cccccccc}
\hline
Galaxy & $S_{60~\micron}$ & $\sigma_{60~\micron}$ & $S_{100~\micron}$ & $\sigma_{100~\micron}$ & $S_{1.4 ~ \rm GHz}$ & $\sigma_{1.4 ~ \rm GHz}$ & $F_{0.1-100 ~ \rm GeV}$\\
 & (Jy) & (mJy) & (Jy) & (mJy) & (mJy) & (mJy) & ($10^{-12}$ erg cm$^{-2}$ s$^{-1}$)\\
\hline
IC 1623 & 22.9 & 62 & 31.6 & 113 & 249 & 9.8 & 0.481\\
NGC 0520 & 31.5 & 30 & 47.4 & 146 & 176 & 5.3 & 0.396\\
NGC 0660 & 65.5 & 88 & 115 & 134 & 373 & 13.9 & 1.06\\
NGC 0695 & 7.59 & 31 & 13.6 & 167 & 74.8 & 3.1 & 1.36\\
NGC 0828 & 11.5 & 53 & 25.3 & 113 & 104 & 3.7 & 0.715\\
MRK 1027 & 5.28 & 39 & 8.57 & 108 & 53.2 & 1.6 & 0.521\\
NGC 0891 & 66.5 & 71 & 172 & 353 & 239 & 7.9 & 2.04\\
NGC 1022 & 19.7 & 35 & 27.3 & 111 & 46.7 & 1.5 & 1.12\\
NGC 1055 & 23.4 & 51 & 65.3 & 82 & 199 & 6.7 & 1.56\\
NGC 1144 & 5.06 & 38 & 11.5 & 198 & 155 & 5.4 & 0.328\\
NGC 1365 & 94.3 & 33 & 166 & 126 & 376 & 13 & 1.33\\
IC 0342 & 181 & 70 & 392 & 434 & 191 & 7.3 & 1.83\\
NGC 1530 & 9.88 & 43 & 25.8 & 128 & 67.3 & 2.7 & 1.77\\
NGC 1614 & 32.1 & 83 & 34.3 & 430 & 137 & 4.9 & 1.14\\
VII Zw 31 & 5.51 & 31 & 10.1 & 126 & 41.4 & 1.3 & 0.151\\
NGC 2276 & 14.3 & 58 & 29.0 & 133 & 267 & 8.6 & 0.522\\
ARP 055 & 6.07 & 48 & 10.3 & 109 & 36.8 & 1.2 & 0.583\\
NGC 2903 & 60.5 & 52 & 130 & 208 & 445 & 13.9 & 1.08\\
UGC 05101 & 11.7 & 34 & 19.9 & 137 & 170 & 5.8 & 0.383\\
NGC 3079 & 50.7 & 64 & 105 & 141 & 769 & 27.1 & 1.40\\
NGC 3147 & 8.17 & 38 & 29.6 & 222 & 89.9 & 3.4 & 0.192\\
ARP 148 & 6.38 & 34 & 10.3 & 106 & 36.4 & 1.2 & 0.797\\
NGC 3556 & 32.6 & 68 & 76.9 & 231 & 217 & 7.1 & 1.11\\
NGC 3627 & 66.3 & 59 & 137 & 118 & 453 & 11.2 & 1.01\\
NGC 3628 & 54.8 & 77 & 106 & 240 & 291 & 8.7 & 0.866\\
ARP 299 & 113 & 52 & 111 & 133 & 677 & 25.4 & 1.08\\
NGC 3893 & 15.6 & 54 & 36.8 & 63 & 139 & 5 & 0.431\\
NGC 4030 & 18.5 & 46 & 50.9 & 117 & 154 & 5.3 & 0.399\\
NGC 4041 & 14.2 & 19 & 31.7 & 126 & 102 & 3.8 & 0.875\\
NGC 4414 & 29.6 & 80 & 70.7 & 72 & 240 & 7.9 & 0.578\\
NGC 4631 & 85.4 & 62 & 160 & 260 & 446 & 14.1 & 0.635\\
MRK 0231 & 30.8 & 42 & 29.7 & 108 & 309 & 12.1 & 0.651\\
NGC 4826 & 36.7 & 77 & 81.7 & 99 & 99.9 & 3.7 & 0.755\\
NGC 5005 & 22.2 & 21 & 63.4 & 131 & 181 & 6.2 & 1.11\\
NGC 5055 & 40.0 & 49 & 140 & 356 & 350 & -- & 0.626\\
ARP 193 & 17.0 & 88 & 24.4 & 120 & 104 & 3.2 & 0.983\\
NGC 5135 & 16.9 & 46 & 31.0 & 177 & 200 & 7.1 & 1.28\\
NGC 5194 & 97.4 & 193 & 221 & 329 & 1570 & -- & 0.723\\
NGC 5236 & 266 & 64 & 524 & 302 & 405 & 15 & 0.609\\
MRK 0273 & 22.5 & 42 & 22.5 & 70 & 145 & 5.1 & 0.449\\
NGC 5678 & 9.67 & 29 & 25.7 & 85 & 110 & 3.9 & 0.496\\
NGC 5713 & 22.1 & 65 & 37.3 & 88 & 158 & 5.7 & 0.203\\
NGC 5775 & 23.6 & 46 & 55.6 & 97 & 280 & 9.1 & 0.680\\
NGC 6240 & 22.9 & 54 & 26.5 & 174 & 426 & 15 & 0.269\\
NGC 6701 & 10.1 & 28 & 20.1 & 67 & 88.9 & 3.1 & 0.426\\
NGC 6946 & 130 & 71 & 291 & 458 & 187 & 6.7 & 0.551\\
IC 5179 & 19.4 & 62 & 37.3 & 90 & 169 & 6.2 & 0.293\\
NGC 7130 & 16.7 & 44 & 25.9 & 132 & 190 & 7.5 & 0.227\\
NGC 7331 & 45.0 & 91 & 110 & 468 & 328 & 8.1 & 1.19\\
NGC 7469 & 27.3 & 40 & 35.2 & 599 & 181 & 5.4 & 0.847\\
NGC 7479 & 14.9 & 49 & 26.7 & 247 & 99.0 & 3.7 & 1.95\\
NGC 7771 & 19.7 & 135 & 40.1 & 839 & 141 & 5.2 & 0.905\\
MRK 0331 & 18 & 46 & 22.7 & 194 & 70.7 & 2.2 & 0.303\\
\hline
\multicolumn{8}{l}{\textit{Notes:} NGC 5194 = M51, NGC 5236 = M83.}\\
\vspace{0.5in}
\label{tab:data2}
\end{tabular}
\end{center}
\end{minipage}
\end{table*}
%
%

\section{Leptonic Emissivities} \label{sec:appb}

\subsection{Bremsstrahlung} \label{sec:brem}

The emissivity for bremsstrahlung is given as
\begin{equation}
q_{\rm Br} = c n_{\rm ISM} \int_{E_{\rm min}}^{\infty} dE_{e} N_{e}(E_{e}) \frac{d\sigma}{dE_{\gamma}},
\end{equation}
where $E_{\rm min} = \sqrt{E_{\gamma} (2 m_{e} c^{2} + E_{\gamma})}$ and the differential cross section is given by \citep{Ginzburg69,Blumenthal70,Stecker71,Schlick02}
\begin{equation}
\frac{d\sigma}{dE_{\gamma}} = \frac{3 \alpha \sigma_{T}}{8 \pi E_{\gamma}} \left[ \left[ 1 + \left( 1 - \frac{E_{\gamma}}{E_{e}} \right)^{2} \right] \phi_{1} + \left( 1 - \frac{E_{\gamma}}{E_{e}} \right) \phi_{2} \right].
\end{equation}
The scattering functions $\phi_{1}$ and $\phi_{2}$ are related such that $\phi_{2} = -2/3 \phi_{1}$ and are equivalent to to $Z^{2} \phi_{u}$ where $\phi_{u}$ is given by \citep{Ginzburg69,Blumenthal70,Schlick02}
\begin{equation}
\phi_{u} = 4 \left[ \text{ln} \left( \frac{2E_{e}}{m_{e}c^{2}} \left( \frac{E_{e} - E_{\gamma}}{E_{\gamma}} \right) \right) - \frac{1}{2}  \right].
\end{equation}
%

\subsection{Photon Number Density} \label{sec:photons}

Throughout the paper, we assume a greybody distribution for optical depths of unity or less.  The photon number density for such a distribution is given by \citep{RL79}
\begin{equation}
n_{ph}(\epsilon) = \frac{C_{\rm dil}}{\pi^{2} \hbar^{3} c^{3}} \frac{\epsilon^{2}}{e^{\epsilon / kT_{\rm rad}} - 1} \left( \frac{\epsilon}{\epsilon_{0}} \right)^{\beta}, \label{eqn:nph}
\end{equation}
where $\beta = 1.6$ and $\lambda_{0} = 200$~\micron, parameters which are adopted from \citet{Casey12}.  The number density is normalized such that \citep[e.g.,][]{Ginzburg69,RL79,Ghisellini13}
\begin{align}
U_{\rm rad} &= \int_{0}^{\infty} \epsilon n_{ph}(\epsilon) d\epsilon \label{eqn:urad} \\
&= \frac{C_{\rm dil}}{\pi^{2} \hbar^{3} c^{3}} \epsilon_{0}^{-\beta} \int_{0}^{\infty} \frac{\epsilon^{3 + \beta}}{e^{\epsilon / kT_{\rm rad}} - 1} d\epsilon \nonumber \\
&= \frac{C_{\rm dil} \epsilon_{0}^{-\beta}}{\pi^{2} \hbar^{3} c^{3}} (kT_{\rm rad})^{4 + \beta} \Gamma (4 + \beta) \zeta (4 + \beta), \nonumber
\end{align}
where $\Gamma(x)$ is the gamma function and $\zeta(x)$ is the Riemann zeta function.  Solving for the dilution factor, $C_{\text{dil}}$, we have
\begin{equation}
C_{\rm dil} = \frac{\pi^{2} \hbar^{3} c^{3} \epsilon_{0}^{\beta} U_{\rm rad}}{(kT_{\rm rad})^{4 + \beta} \Gamma ( 4 + \beta) \zeta ( 4 + \beta)}, \label{eqn:cdil}
\end{equation}
and substituting back into the number density gives us,
\begin{equation}
n_{ph}(\epsilon) = \frac{U_{\rm rad}}{(kT_{\rm rad})^{4+\beta} \Gamma(4+\beta) \zeta(4 + \beta)} \times \frac{\epsilon^{2 + \beta}}{e^{\epsilon / kT_{\rm rad}} - 1}. \label{eqn:nph_full}
\end{equation}
In the optically thin case, $\beta = 1.6$ and thus $\Gamma (5.6) \zeta(5.6) \approx 63$.  In the optically thick case, $\beta = 0$ and thus $\Gamma (4) \zeta(4) = \pi^{2} / 15 \approx 6.5$.

\subsection{Inverse Compton Emissivity} \label{sec:ic}

The emissivity for inverse Compton emission is given as \citep{Blumenthal70,Schlick02}
\begin{equation}
q_{\rm IC}(E_{\gamma}) = \frac{3}{4} c \sigma_{T} \int_{0}^{\infty} d\epsilon \frac{n_{ph}(\epsilon)}{\epsilon} \int_{E_{\rm min}}^{\infty} dE_{e} \frac{N_{e}(E_{e})}{\gamma_{e}^{2}} F(q, \Gamma), \label{eqn:qic}
\end{equation}
with units of GeV$^{-1}$~cm$^{-3}$~s$^{-1}$, where the minimum cosmic ray electron energy is given by
\begin{equation}
E_{\rm min} = \frac{1}{2} E_{\gamma} \left[ 1 + \left( 1 + \frac{m_{e}^{2}c^{4}}{\epsilon E_{\gamma}} \right)^{1/2} \right]. \nonumber
\end{equation}
Here, $E_{\gamma}$ is the energy of the resulting $\gamma$-ray and $\epsilon$ is the energy of the incident photon.  The function $F(q, \Gamma)$ is part of the Klein-Nishina cross section and is given by \citep{Ginzburg69,Blumenthal70,Stecker71,RL79,Schlick02,Dermer09,Bottcher12,Ghisellini13} 
\begin{equation}
F(q, \Gamma) = 2q \ln(q) + (1 + q - 2q^{2}) + \frac{\Gamma^{2}q^{2} (1 - q)}{2 (1 + \Gamma q)}, \label{eqn:fic}
\end{equation}
where
\begin{equation}
\Gamma = \frac{4 \epsilon \gamma_{e}}{m_{e}c^{2}} ~~ \text{and} ~~ q = \frac{E_{\gamma}}{\Gamma (\gamma_{e} m_{e} c^{2} - E_{\gamma})}. \nonumber
\end{equation}
Here, $\Gamma$ is a product of the photon and electron energies and is not to be confused with the gamma function, $\Gamma(x)$, or the cosmic ray spectral index, $\Gamma_{p}$.

To better understand the effects of the dust temperature ($T_{\rm rad}$) and to simplify our expression for the inverse Compton emissivity, we start by designating the second integral as $I(\epsilon, E_{\gamma})$ such that
\begin{equation}
q_{\rm IC}(E_{\gamma}) = \frac{3}{4} c \sigma_{T} \int_{0}^{\infty} d\epsilon \frac{n_{ph}(\epsilon)}{\epsilon} I(\epsilon, E_{\gamma}), \label{eqn:qic2}
\end{equation}
where
\begin{equation}
I(\epsilon, E_{\gamma}) = \int_{E_{\rm min}}^{\infty} dE_{e} \frac{N_{e}(E_{e})}{\gamma_{e}^{2}} F(q, \Gamma). \label{eqn:int}
\end{equation}
Then, substituting our expression for the greybody distribution from above into our expression for emissivity, we have
\begin{equation*}
q_{\rm IC}(E_{\gamma}) = \frac{3 c \sigma_{T} U_{\rm rad}}{4 K(\beta)(kT_{\rm rad})^{4 + \beta}} \int_{0}^{\infty} d\epsilon \frac{\epsilon^{1 + \beta}}{e^{\epsilon / kT_{\rm rad}} - 1} I(\epsilon, E_{\gamma}), \label{eqn:qic3}
\end{equation*}
where $K(\beta) = \Gamma(4 + \beta) \zeta(4 + \beta)$.  Changing variables from $\epsilon$ to $u = \epsilon / kT_{\rm rad}$ gives
\begin{align}
q_{\rm IC}(E_{\gamma}) &= \frac{3 c \sigma_{T} U_{\rm rad}}{4 K(\beta)(kT_{\rm rad})^{4 + \beta}} \int_{0}^{\infty} du \times (kT_{\rm rad}) \nonumber \\
&\times \frac{u^{1 + \beta} \times (kT_{\rm rad})^{1 + \beta}}{e^{u} - 1} I(u \times (kT_{\rm rad}), E_{\gamma}) \nonumber \\
q_{\rm IC}(E_{\gamma}) &= \frac{3 c \sigma_{T} U_{\rm rad}}{4 K(\beta)(kT_{\rm rad})^{2}} \int_{0}^{\infty} \frac{u^{1 + \beta} du}{e^{u} - 1} I(u \times kT_{\rm rad}, E_{\gamma}). \label{eqn:qic_fin}
\end{align}
Thus, the inverse Compton emissivity decreases as the square of the dust temperature and also changes as $I(\epsilon, E_{\gamma})$ changes.  As $I(\epsilon, E_{\gamma})$ depends on $N_{e}(E_{e}) = q_{e}(E_{e}) t_{e}(E_{e})$, the expression will change as the electron lifetime changes with dust temperature.

\section{Photon Interactions} \label{sec:appc}

For photon-photon ($\gamma-\gamma$) interactions, the cross section is given by
\begin{align}
\sigma_{\gamma-\gamma}(\beta_{\rm cm}) &= \frac{3}{16} \sigma_{T} (1 - \beta_{\rm cm}^{2}) \nonumber \\
&\times \left[ (3 - \beta_{\rm cm}^{4}) \ln \left( \frac{1 + \beta_{\rm cm}}{1 - \beta_{\rm cm}} \right) - 2 \beta_{\rm cm} (2 - \beta_{\rm cm}^{2}) \right],
\end{align}
where $\beta_{\rm cm} = (1 - \gamma_{\rm cm}^{-2})^{1/2}$ and $\gamma_{\rm cm}$ is the center-of-momentum frame Lorentz factor of the produced electron and positron \citep{Dermer09,Bottcher12}.  Then, the optical depth for $\gamma-\gamma$ interactions is
\begin{equation} \label{eqn:tau}
\frac{d\tau_{\gamma-\gamma}}{dx}(\epsilon_{1}) = \frac{2}{\epsilon_{1}} \int_{1/\epsilon_{1}}^{\infty} d\epsilon \frac{n_{ph}(\epsilon)}{\epsilon^{2}} \int_{0}^{s_{0}} ds s \sigma_{\gamma-\gamma}(s),
\end{equation}
where $s_{0} = \epsilon\epsilon_{1}$ is the maximum center-of-momentum energy for head-on collisions.  \citet{Dermer09} define $d\tau_{\gamma-\gamma}/dx$ as the absorption probability per unit path length,  so the approximate optical depth is merely $\tau_{\gamma-\gamma} \simeq l_{\rm path} \times d\tau_{\gamma-\gamma}/dx$.  For an inclined ellipse, we take the path length through the disc to be $l_{\rm path} = \pi/2 \times R ~ \sec (90^{\circ} - i)$.

If we substitute into equation~\ref{eqn:tau} with our expression for the greybody distribution from Section~\ref{sec:photons} and change variables from $\epsilon$ to $u = \epsilon/kT_{\rm rad}$, we have
\begin{align}
\frac{d\tau_{\gamma-\gamma}}{dx}(\epsilon_{1}) &= \frac{2}{\epsilon_{1}} \times \frac{3 c \sigma_{T} U_{\rm rad}}{4 K(\beta) (kT_{\rm rad})^{4 + \beta}} \int_{u_{\rm min}}^{\infty} du \times(kT_{\rm rad}) \nonumber \\
&\times \frac{u^{\beta} \times (kT_{\rm rad})^{\beta}}{e^{u} - 1} \int_{0}^{s_{0}} ds s \sigma_{\gamma-\gamma}(s),\\
\frac{d\tau_{\gamma-\gamma}}{dx}(\epsilon_{1}) &= \frac{2}{\epsilon_{1}} \times \frac{3 c \sigma_{T} U_{\rm rad}}{4 K(\beta) (kT_{\rm rad})^{3}} \int_{u_{\rm min}}^{\infty} du \frac{u^{\beta}}{e^{u} - 1} \nonumber \\
&\times \int_{0}^{s_{0}} ds s \sigma_{\gamma-\gamma}(s).
\end{align}
Thus, for $\gamma-\gamma$ interactions, the optical depth is directly proportional to $U_{\rm rad} / (kT_{\rm rad})^{3}$.

For proton-photon ($p-\gamma$) interactions, we utilize the parameterizations presented in \citet{Hummer10}, which is in turn based on the \textsc{SOFIA} code \citep{Mucke00}.  Their parameterizations cover photopion production through both direct and multi-photon production and via the $\Delta$-resonance and higher order $\Delta$- and $N$- resonances.  The parameterized source functions (equation 28) take the form
\begin{equation}
Q_{b}^{\rm IT} = N_{p} \left( \frac{E_{b}}{\chi^{\rm IT}} \right) \frac{m_{p}}{E_{b}} \int_{\epsilon_{\rm th} / 2}^{\infty} dy n_{\gamma} \left( \frac{m_{p} y \chi^{\rm IT}}{E_{b}} \right) M_{b}^{\rm IT} f^{\rm IT}(y),
\end{equation}
where $y = E_{p} \epsilon / m_{p}$ and $\chi, M$ are parameterization constants given in Tables 4-6  in \citet{Hummer10}.  The functions $f(y)$ are further parameterizations given in their equations 31-35 and 38-39.  The parameterization functions are given for each of the types of production (resonances, direct, multi-pion) and the parameterization constants are further broken down for each type of pion ($\pi^{+}, \pi^{-}, \pi^{0}$).

\end{document}